\documentclass[12pt]{article}
\def\paradot#1{\vspace{1.3ex plus 0.7ex minus 0.5ex}\noindent{\bf\boldmath{#1.}}}
\usepackage{latexsym,amsmath,amssymb,amsthm,amsfonts,graphicx,natbib}
\include{def}
\def\,{\mskip 3mu} \def\>{\mskip 4mu plus 2mu minus 4mu} \def\;{\mskip 5mu plus 5mu} \def\!{\mskip-3mu}
\def\dispmuskip{\thinmuskip= 3mu plus 0mu minus 2mu \medmuskip=  4mu plus 2mu minus 2mu \thickmuskip=5mu plus 5mu minus 2mu}
\def\textmuskip{\thinmuskip= 0mu                    \medmuskip=  1mu plus 1mu minus 1mu \thickmuskip=2mu plus 3mu minus 1mu}
\textmuskip
\def\beq{\dispmuskip\begin{equation}}    \def\eeq{\end{equation}\textmuskip}
\def\beqn{\dispmuskip\begin{displaymath}}\def\eeqn{\end{displaymath}\textmuskip}
\def\bea{\dispmuskip\begin{eqnarray}}    \def\eea{\end{eqnarray}\textmuskip}
\def\bqan{\dispmuskip\begin{eqnarray*}}  \def\eqan{\end{eqnarray*}\textmuskip}

\renewcommand{\baselinestretch}{1.6}

\def\G{\Gamma}
\def\a{\alpha}

\def\s{\sigma}

\def\b{\beta}

\def\tr{\text{\rm tr}}

\def\argmin{\text{\rm argmin}}
\def\argmax{\text{\rm argmax}}

\def\diag{\text{\rm diag}}

\begin{document}

\title{Variational approximation for heteroscedastic linear \\
models and matching pursuit algorithms}
\author{David J. Nott, Minh-Ngoc Tran, Chenlei Leng\footnote{David J. Nott and Chenlei Leng are Associate Professors, 
Minh-Ngoc Tran is PhD candidate at
Department of Statistics and Applied Probability,
National University of Singapore, Singapore 117546, Republic of Singapore.
Corresponding author: Minh-Ngoc Tran (ngoctm@nus.edu.sg).}}  
\date{\empty}
\maketitle
\renewcommand{\baselinestretch}{1.2}
\begin{abstract}
Modern statistical applications involving large data sets have focused attention
on statistical methodologies which are both efficient computationally and able to deal with the screening of large
numbers of different candidate models.  Here we consider computationally efficient variational Bayes
approaches to inference in high-dimensional 
heteroscedastic linear regression,
where both the mean and variance are described in terms of linear functions of the predictors and where the number
of predictors can be larger than the sample size.  
We derive a closed form variational lower bound on the log marginal likelihood 
useful for model selection, and propose a novel fast greedy search 
algorithm on the model space which makes use of one-step optimization updates to the variational lower
bound in the current model for screening large numbers of candidate predictor variables for inclusion/exclusion in 
a computationally thrifty way.  We show that the model search strategy we suggest is 
related to widely used orthogonal matching pursuit algorithms for model search but 
yields a framework for potentially extending
these algorithms to more complex models.
The methodology is applied in simulations and in two real examples 
involving prediction for food constituents using NIR technology and prediction of disease progression
in diabetes.

\paradot{Keywords}
Bayesian model selection, 
heteroscedasticity, 
matching pursuit, 
variational approximation.
\end{abstract}  

\newpage 

\section{Introduction}
Consider the heteroscedastic linear regression model
\beq\label{mainmodel}
y_i=x_i^T\beta+\sigma_i\epsilon_i,\ i=1,\dots, n
\eeq
where $y_i$ is a response, $x_i=(x_{i1},...,x_{ip})^T$
is a corresponding $p$-vector of predictors, $\beta=(\beta_1,...,\beta_p)^T$
is a vector of unknown mean parameters, $\epsilon_i\sim N(0,1)$ are
independent errors and
\beqn
\log \sigma_i^2=z_i^T\alpha,
\eeqn
where $z_i=(z_{i1},...,z_{iq})^T$ is a $q$-vector of predictors and
$\alpha=(\alpha_1,...,\alpha_q)^T$ are unknown variance parameters.  
In this model the standard deviation $\sigma_i$ of $y_i$ is being
modeled in terms of the predictors $z_i$; 
this heteroscedastic model may be contrasted with the usual
homoscedastic model which assumes $\sigma_i$ is constant.
We take a Bayesian approach
to inference for this model and consider a prior distribution $p(\theta)$
on $\theta=(\beta^T,\alpha^T)^T$ of the form $p(\theta)=p(\beta)p(\alpha)$ with
$p(\beta)$ and $p(\alpha)$ both normal, $N(\mu_{\beta}^0,\Sigma_{\beta}^0)$
and $N(\mu_{\alpha}^0,\Sigma_{\alpha}^0)$ respectively.
It is possible to consider hierarchical extensions for the priors on 
$p(\beta)$ and $p(\alpha)$ but we do not consider this here.

We will consider a variational Bayes approach for inference (see Section \ref{Max_lb} for the details).
The term variational approximation refers to a wide
range of different methods where the idea is to convert a problem of integration
into an optimization problem.
In Bayesian inference, variational approximation provides a fast alternative to Monte Carlo methods
for approximating posterior distributions in complex models, especially in high-dimensional problems. 
In the heteroscedastic linear regression model, we 
will consider a variational approximation to the joint posterior
distribution of $\beta$ and $\alpha$ as $q(\beta,\alpha)=q(\beta)q(\alpha)$,
where $q(\beta)$ and $q(\alpha)$ are both normal densities, 
$N(\mu_{\beta}^q,\Sigma_{\beta}^q)$ and $N(\mu_{\alpha}^q,\Sigma_{\alpha}^q)$ respectively.  
It is also possible to give a variational treatment in which independence is not assumed
between $\beta$ and $\alpha$ (John Ormerod, personal communication) 
but this complicates the variational optimization somewhat.  
We attempt to choose the parameters in the variational posterior $\mu_{\beta}^q$, 
$\mu_{\alpha}^q$, $\Sigma_{\beta}^q$ and $\Sigma_{\alpha}^q$ to minimize the Kullback-Leibler
divergence between the true posterior distribution $p(\beta,\alpha|y)$ and $q(\beta,\alpha)$. 
This results in a lower bound on the log marginal likelihood $\log p(y)$ - a key quantity in Bayesian model selection.
The first contribution of our paper is the derivation of a closed form for the lower bound
and the proposal of an iterative scheme for maximizing it.
This lower bound maximization plays a crucial role in the variable selection problem
discussed in Section \ref{modelselection}.  

Variable selection is a fundamental problem in statistics and machine learning,
and a large number of methods have been proposed for variable selection in homoscedastic regression.
The traditional variable selection approach in the Bayesian framework consists of
building a hierarchical Bayes model and using MCMC algorithms to estimate posterior model probabilities
\citep{George:1993,Smith:1996,Raftery:1997}.
This methodology is computationally demanding in high-dimensional problems
and there is a need for fast alternatives in some applications.
In high-dimensional settings, alternative approaches include 
the family of greedy algorithms \citep{Tropp:2004,Zhang:2009},
also known as {\em matching pursuit} \citep{Mallat:1993} in signal processing.
Greedy algorithms are closely related to the Lasso \citep{Tibshirani:1996} and the LARS algorithm \citep{Efron:2004}.
See \cite{Zhao:2007,Efron:2004} and \cite{Zhang:2009} for excellent comparisons of these families of algorithms.
In the statistical context, greedy algorithms have been proven to be
very efficient for variable selection in linear regression 
under the assumption of homoscedasticity, i.e. where the variance is assumed to be constant \citep{Zhang:2009}.

In many applications the assumption of constant variance may be unrealistic.
Ignoring heteroscedasticity may lead to serious problems in inference, such as misleading
assessments of significance, poor predictive performance and inefficient estimation of
mean parameters.  In some cases, learning the structure in the variance may be the primary goal.
See \cite{Chan:2006}, \cite{Carroll:1988} and 
\cite{Ruppert:2003}, Chapter 14, for a more detailed discussion on heteroscedastic modeling.
Despite a large number of works on heteroscedastic linear regression and 
overdispersed generalized linear models in which overdispersion
is modeled to depend on the covariates 
\citep{Efron:1986,Nelder:1987,Davidian:1987,Smyth:1989,Yee:1996,Rigby:2005},
methods for variable selection seem to be somewhat overlooked.  
\cite{Yau:2003} and \cite{Chan:2006} consider Bayesian variable selection using MCMC computational approaches 
in heteroscedastic Gaussian models.
They discuss extensions involving flexible modeling of the mean and
variance functions.  \cite{Cottet:2008} consider extensions to overdispersed generalized linear
and generalized additive models.  These approaches are computationally demanding in high-dimensional
settings.  A general and flexible framework for modeling overdispersed data is also considered
by \cite{Yee:1996} and \cite{Rigby:2005}.  
Methods for model selection, however, are less well developed.
A common approach is to use information criteria such as generalized AIC and BIC
together with forward stepwise methods (see, for example, \cite{Rigby:2005}, Section 6).  
We compare our own approaches to such methods later.  
A main contribution of the present paper is to propose
a novel fast greedy algorithm for variable selection in heteroscedastic linear regression.
We show that the proposed algorithm is in homoscedastic cases similar to currently used methods
while having many attractive properties and working efficiently in high-dimensional problems.
An efficient R program is available on the authors' websites.

\begin{figure*}[h]
\centerline{\includegraphics[width=1\textwidth,height=.7\textwidth]{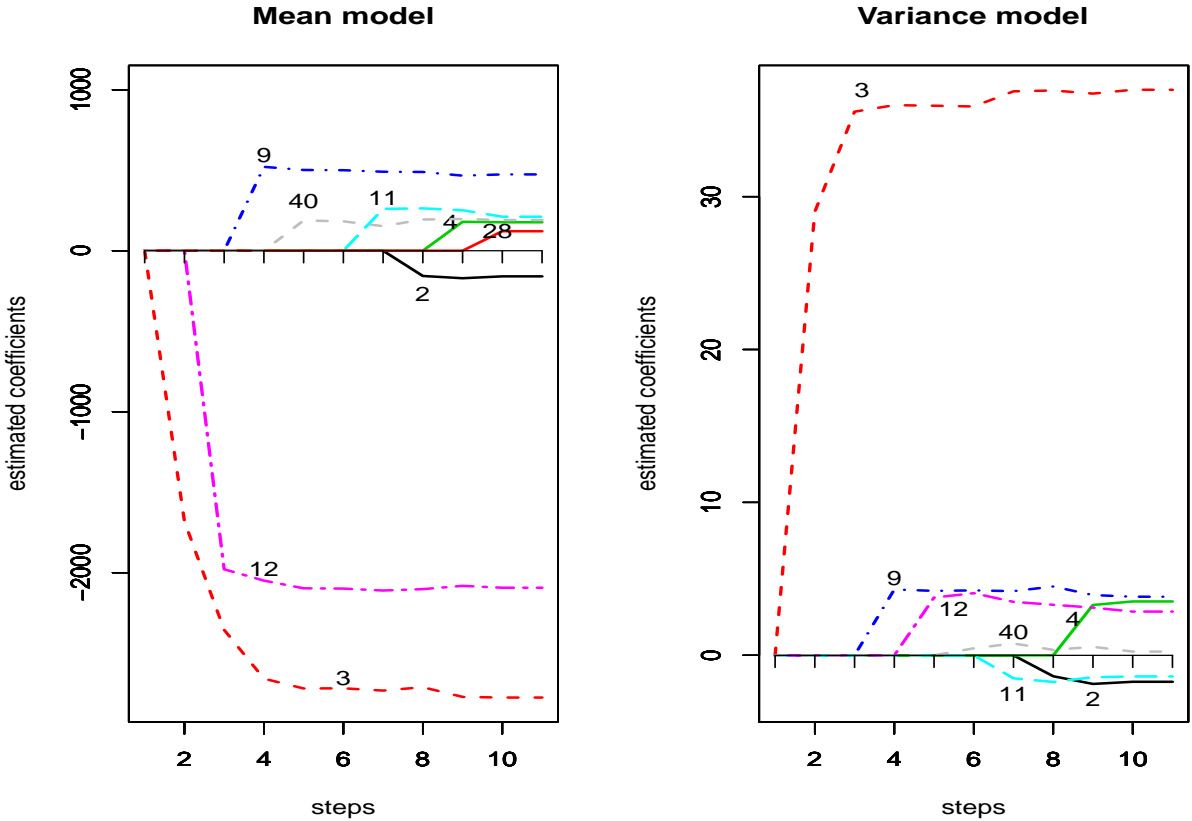}}
\caption{\label{solutionpath}
Solution paths as functions of iteration steps for analyzing the diabetes data
using heteroscedastic linear regression. The algorithm stops after 11 iterations 
with 8 and 7 predictors selected for the mean and variance models respectively.
The selected predictors enter the mean (variance) model in the order 3, 12, ..., 28 (3, 9, ..., 4).}
\end{figure*}

In Section \ref{secApp} we apply our algorithm to the analysis of the diabetes data \citep{Efron:2004}
using heteroscedastic linear regression. 
This data set consists of 64 predictors (constructed from 10 input variables for a ``quadratic model") and 442 observations. 
We show in Figure \ref{solutionpath} the estimated coefficients corresponding to 
selected predictors as functions of iteration steps in our algorithm,
for both the mean and variance models.
The algorithm stops after 11 forward selection steps
with 8 and 7 predictors selected for the mean and variance models respectively.

The rest of the paper is organized as follows.
The closed form of the lower bound and the iterative scheme for maximizing it
are presented in Section \ref{Max_lb}.
We present in Section \ref{modelselection} our novel fast greedy algorithm,
and compare it to existing greedy algorithms in the literature for homoscedastic regression.
In Section \ref{secApp} we apply our methodology 
to the analysis of two benchmark data sets. 
Experimental studies are presented in Section \ref{secSimul}.
Section \ref{discussion} contains conclusions and future research.     
Technical derivation is relegated to the Appendices.
\section{Variational Bayes}\label{Max_lb}
We now give a brief introduction to the variational approximation method. 
For a more detailed exposition see, for example, 
\cite{Jordan:1999}, \cite{Bishop:2006} Chapter 10,
or see \cite{Ormerod:2009} for a statistically oriented introduction.  
The term variational approximation refers to a wide
range of different methods where the idea is to convert a problem of integration
into an optimization problem.  Here we will only be concerned with applications
of variational methods in Bayesian inference and only with a particular approach
sometimes referred to as parametric variational approximation.  
Write $\theta$ for all our unknown parameters, $p(\theta)$ for
the prior distribution and $p(y|\theta)$ for the likelihood.  
For Bayesian inference, decisions are based on the posterior distribution $p(\theta|y)\propto p(\theta)p(y|\theta)$.
A common difficulty in applications is how to compute quantities
of interest with respect to the posterior. 
These computations often involve the evaluation
of high-dimensional integrals.  Variational approximation proceeds by approximating
the posterior distribution directly.  Formally, we consider a family of distributions 
$q(\theta|\lambda)$ where $\lambda$ denotes some unknown parameters, and attempt
to choose $\lambda$ so that $q(\theta|\lambda)$ is closest to $p(\theta|y)$ in some
sense.  In particular, we attempt to minimize the Kullback-Leibler divergence
\beqn\int \log \frac{q(\theta|\lambda)}{p(\theta|y)}q(\theta|\lambda) d\theta\eeqn
with respect to $\lambda$.  Using the identity
\begin{eqnarray}
 \log p(y) & = & \int \log \frac{p(\theta)p(y|\theta)}{q(\theta|\lambda)}q(\theta|\lambda)d\theta
+\int \log \frac{q(\theta|\lambda)}{p(\theta|y)}q(\theta|\lambda) d\theta, \label{keyidentity}
\end{eqnarray}
where $p(y)=\int p(\theta)p(y|\theta)d\theta$, we see that minimizing the Kullback-Leibler divergence
is equivalent to the maximization of 
\begin{eqnarray}
 & & \int \log \frac{p(\theta)p(y|\theta)}{q(\theta|\lambda)}q(\theta|\lambda)d\theta.   \label{lowerbd}
\end{eqnarray}
Here \eqref{lowerbd} is a lower bound (often called the free energy in physics)
on the log marginal likelihood $\log p(y)$ due to the non-negativity
of the Kullback-Leibler divergence term in (\ref{keyidentity}).  
The lower bound (\ref{lowerbd}), when maximized with respect
to $\lambda$, is often used as an approximation to the log marginal likelihood $\log p(y)$.
The error in the approximation is the 
Kullback-Leibler divergence between the approximation $q(\theta|\lambda)$ and the true
posterior.  The approximation
is useful, since $\log p(y)$ is a key quantity in Bayesian model selection.  

For our heteroscedastic linear model, the lower bound (\ref{lowerbd}) can be expressed as 
\beqn
L=T_1+T_2+T_3,
\eeqn
where
\bqan
T_1 &=& \int \log [p(\beta,\alpha)] q(\beta)q(\alpha) d\beta d\alpha,\\
T_2 &=&\int \log [p(y|\beta,\alpha)] q(\beta) q(\alpha) d\beta d\alpha,\\
T_3 &=& -\int \log \left[ q(\beta)q(\alpha)\right] q(\beta)q(\alpha) d\beta d\alpha.
\eqan
We show (see the Appendix A) that these three terms, which are all expectations with respect to the (assumed normal)
variational posterior, can be evaluated analytically.  Putting the terms together we obtain
that the lower bound (\ref{lowerbd}) on the log marginal likelihood is
\bea
L & = & \frac{p+q}{2}-\frac{n}{2}\log 2\pi+\frac{1}{2}\log |\Sigma_{\beta}^q {\Sigma_{\beta}^0}^{-1}|+\frac{1}{2}\log |\Sigma_{\alpha}^q {\Sigma_{\alpha}^0}^{-1}|-\frac{1}{2}\tr ({\Sigma_\beta^0}^{-1} \Sigma_\beta^q)\nonumber\\
&& -\frac{1}{2}\tr ({\Sigma_\alpha^0}^{-1} \Sigma_\alpha^q)-\frac{1}{2}(\mu_{\beta}^q-\mu_{\beta}^0)^T {\Sigma_{\beta}^0}^{-1} (\mu_{\beta}^q-\mu_{\beta}^0)-\frac{1}{2}(\mu_{\alpha}^q-\mu_{\alpha}^0)^T {\Sigma_{\alpha}^0}^{-1} (\mu_{\alpha}^q-\mu_{\alpha}^0) \nonumber \\
&& -\frac{1}{2}\sum_{i=1}^n z_i^T\mu_{\alpha}^q -\frac{1}{2}\sum_{i=1}^n \frac{(y_i-x_i^T\mu_\beta^q)^2+x_i^T\Sigma_{\beta}^q x_i}
{\exp\left(z_i^T\mu_{\alpha}^q-\frac{1}{2}z_i^T\Sigma_{\alpha}^q z_i\right)}. \label{hetlowerbd}
\eea
This needs to be maximized with respect to $\mu_{\beta}^q$, $\mu_{\alpha}^q$,
$\Sigma_{\beta}^q$, $\Sigma_{\alpha}^q$.  We consider an
iterative scheme in which we maximize with respect to each of the blocks of parameters $\mu_\beta^q$, $\mu_{\alpha}^q$,
$\Sigma_{\beta}^q$, $\Sigma_{\alpha}^q$ with the other blocks held fixed.  

Write $X$ for the design matrix with $i$th row $x_i^T$ and $D$ for the diagonal matrix
with $i$th diagonal element $1/\exp(z_i^T\mu_\alpha^q-\frac12z_i^T\Sigma_\alpha^qz_i)$.  
Maximization with respect to $\mu_\beta^q$ with other terms held fixed leads to
\beqn
\mu_\beta^q=\left(X^TDX+{\Sigma_{\beta}^0}^{-1}\right)^{-1}\left({\Sigma_{\beta}^0}^{-1}\mu_\beta^0+X^T Dy\right).
\eeqn
Maximization with respect to $\Sigma_{\beta}^q$ with other terms held fixed leads to
\beqn
\Sigma_\beta^q=\left({\Sigma_{\beta}^0}^{-1}+X^TDX\right)^{-1}.
\eeqn

Handling the parameters $\mu_\alpha^q$ and $\Sigma_\alpha^q$ 
in the variational posterior for $\alpha$ is more complex.  We proceed in the following way.  
If no parametric form for the variational posterior $q(\alpha)$ is assumed (that is, if we
do not assume that $q(\alpha)$ is normal) but only assume the factorization $q(\theta)=q(\beta)q(\alpha)$
then the optimal choice for $q(\alpha)$ for a given $q(\beta)=N(\mu_\beta^q,\Sigma_\beta^q)$ 
is (see \cite{Ormerod:2009}, for example)
\beq
q(\alpha)\propto \exp\Big(E(\log [p(\theta)p(y|\theta)])\Big),\label{optdis1}
\eeq
where the expectation is with respect to $q(\beta)$.  
Similar to the derivation of the lower bound \eqref{hetlowerbd},
it is easy to see that
\beqn
q(\a)\propto \exp\left(-\frac12\sum_{i=1}^nz_i^T\a-\frac12\sum_{i=1}^n\frac{(y_i-x_i^T\mu_\beta^q)^2+x_i^T\Sigma_\beta^qx_i}{\exp(z_i^T\a)}
-\frac12(\a-\mu_\a^0)^T{\Sigma_\a^0}^{-1}(\a-\mu_\a^0)\right),
\eeqn
which takes the form of the 
posterior (apart from a normalization constant)
for a Bayesian generalized linear model with gamma response and log link, coefficient of variation $\sqrt{2}$, 
and responses $w_i=(y_i-x_i^T\mu_\beta^q)^2+x_i^T\Sigma_\beta^q x_i$ with 
the log of the mean response being $z_i^T\alpha$.  The prior in this gamma generalized linear model
is $N(\mu_\alpha^0,\Sigma_\alpha^0)$.  If we use a quadratic approximation to $\log q(\a)$
then this results in a normal approximation to $q(\alpha)$.  
We choose the mean and variance of the normal approximation simply by the posterior mode 
and the negative inverse Hessian 
of the log posterior at the mode for the gamma generalized linear model described above.  
The computations required are standard ones
in fitting a Bayesian generalized linear model (see Appendix B).  
Write $Z$ for the design matrix in the variance model with $i$th row $z_i^T$ and 
write $W(\a)$ (as a function of $\a$) for the diagonal matrix $\diag(\frac12w_i\exp(-z_i^T\a))$.
With $\mu_\alpha^q$ the posterior mode, we obtain for $\Sigma_\alpha^q$
the expression
\beqn
\Sigma_{\alpha}^q=\left(Z^T W(\mu_\a^q) Z+{\Sigma_{\alpha}^0}^{-1}\right)^{-1}.
\eeqn 
Our optimization over $\mu_\alpha^q$ and $\Sigma_\alpha^q$ is only approximate, so that
we only retain the new values in the optimization if they result in an improvement in the
lower bound (\ref{hetlowerbd}).  The advantage of our approximate approach is the closed form
expression for the update of $\Sigma_\alpha^q$ once $\mu_\alpha^q$ is found, so that explicit
numerical optimization for a possibly high-dimensional covariance matrix is avoided.  

The explicit algorithm for our method is the following.  

\paradot{Algorithm 1: Maximization of the variational lower bound}
\begin{enumerate}
\item Initialize parameters $\mu_\a^q,\ \Sigma_\a^q$. 
\item $\mu_{\beta}^q \leftarrow \left(X^TDX+{\Sigma_{\beta}^0}^{-1}\right)^{-1}
\left({\Sigma_{\beta}^0}^{-1}\mu_{\beta}^0+X^TD y\right)$ where $D$ is
the diagonal matrix with $i$th diagonal entry 
$1/\exp\left(z_i^T\mu_{\alpha}^q-1/2z_i^T\Sigma_{\alpha}^q z_i\right)$.  
\item $\Sigma_{\beta}^q \leftarrow \left(X^T D X+{\Sigma_{\beta}^0}^{-1}\right)^{-1}$. 
\item Obtain $\mu_{\alpha}^q$ as the posterior mode for a gamma generalized linear model with
normal prior $N(\mu_{\alpha}^0,\Sigma_{\alpha}^0)$, 
gamma responses $w_{i}=(y_i-x_i^T\mu_{\beta}^q)^2+x_i^T\Sigma_{\beta}^q x_i$, 
coefficient of variation $\sqrt{2}$ and where the log of the mean is 
$z_i^T\alpha$.  
\item $\Sigma_{\alpha}^q\leftarrow \left(Z^T W Z+{\Sigma_{\alpha}^0}^{-1}\right)^{-1}$ 
where $W$ is diagonal with $i$th diagonal element $w_{i}\exp(-z_i^T\mu_{\alpha }^q)/2$.  
\item If the updates done in steps 3 and 4 do not improve the lower bound (\ref{hetlowerbd}) 
then their old values are retained.   
\item Repeat steps 2-6 until the increase in the variational lower bound \eqref{hetlowerbd}
is less than some user specified tolerance.  
\end{enumerate}
For initialization, 
we first perform an ordinary least squares (OLS) fit for the mean model to get an estimate $\hat{\beta}$ of
$\beta$.  Then we take the residuals from this fit, say $r_i=(y_i-x_i^T\hat{\beta})^2$, and do an OLS fit
of $\log r_i$ to the predictors $z_i$ to obtain our initial estimate 
of $\mu_\alpha^q$.  The initial value of $\Sigma_\alpha^q$ is then set to the covariance matrix
of the least squares estimator.  
When the OLS fits are not valid, some other method such as the Lasso
can be used instead.
The application of this algorithm to the problem of variable selection in Section \ref{modelselection}
always involves only situations in which the above OLS fits are available.

We mention one further extension of our method.  We have assumed above that the prior covariance matrices
$\Sigma_\beta^0$ and $\Sigma_\alpha^0$ are known.  Later we will assume $\Sigma_\beta^0=\sigma_\beta^2I$
and $\Sigma_\alpha^0=\sigma_\alpha^2 I$ where $I$ denotes the identity matrix and $\sigma_\beta^2$ and
$\sigma_\alpha^2$ are scalar variance parameters.  
We further assume that $\mu_\beta^0=0$ and $\mu_\alpha^0=0$.
It may be helpful to perform some data driven shrinkage
so that $\sigma_\beta^2$ and $\sigma_\alpha^2$ are considered unknown and to be estimated from the data.  
Our lower bound (\ref{hetlowerbd}) can be considered as an approximation to $\log p(y|\sigma_\beta^2,\sigma_\alpha^2)$,
and the log posterior for $\sigma_\beta^2,\sigma_\alpha^2$ is apart from an additive constant
\beqn\log p(\sigma_\beta^2,\sigma_\alpha^2)+\log p(y|\sigma_\beta^2,\sigma_\alpha^2).\eeqn
If we assume independent inverse gamma priors, $IG(a,b)$, for $\sigma_\beta^2$ and $\sigma_\alpha^2$, 
and if we replace the log marginal likelihood by the lower bound and maximize, we get
\beqn
\sigma_\beta^2=\frac{b+\frac12 {\mu_\beta^q}^T \mu_\beta^q+\frac12\tr(\Sigma_\b^q)}{a+1+p/2}
\eeqn
and
\beqn
\sigma_\alpha^2=\frac{b+\frac12{\mu_\alpha^q}^T \mu_\alpha^q+\frac12\tr(\Sigma_\a^q)}{a+1+q/2}.
\eeqn
These updating steps can be added to the Algorithm 1 given above.   

\section{Model selection}\label{modelselection}
In the discussion of the previous sections the choice of predictors in the mean and variance models
was fixed.  We now consider the problem of variable selection in the heteroscedastic
linear model, and the question of computationally efficient
model search when the number of candidate predictors is very large, perhaps much larger
than the sample size.  In Sections 1 and 2 we denoted the marginal likelihood by $p(y)$ without
making explicit conditioning on the model but now we write $p(y|m)$ for the marginal likelihood
in a model $m$.  If we have a prior distribution $p(m)$ on the set of all models under consideration
then Bayes' rule leads to the posterior distribution on the model given by
$p(m|y)\propto p(m)p(y|m)$.  We can use the variational lower bound on $\log p(y|m)$ 
as a replacement
for $\log p(y|m)$ in this formula as one strategy for Bayesian variable selection when $p(y|m)$ is
difficult to compute. We follow that strategy here.  
For a more thorough review of the Bayesian
approach to model selection see, for example, \cite{O'Hagan:2004}.

Using the maximized lower bound is a popular approach for model selection
\citep{Beal:2003,Beal:2006,Wu:2010}.
The error in using the lower bound to approximate $\log p(y|m)$
is the Kullback-Leibler divergence between the true posterior $p(\theta|y)$
and the variational distribution $q(\theta|\lambda)$. 
The true posterior in our model has the structure of a product of two normal distributions
and the variational distribution we use is also a product of two normals. 
Therefore, it can be expected that the minimized KL divergence is small, 
thus leading to a good approximation. 
The experimental study in Section 5 suggests that the maximized lower bound is very tight. 

Before presenting our strategy for ranking variational lower bounds,
we discuss here the model prior.
Suppose we have a current model with predictors $x_i$, $i\in C\subset D=\{1,...,p\}$ in the mean model
and $z_i$, $i\in V\subset E=\{1,...,q\}$ in the variance model.  
The subsets $C$ and $V$ give indices for the currently active predictors
in the mean and variance models.  
Let $\pi^\mu_i$ ($\pi^\s_j$) be the prior probability for inclusion of $x_i$ ($z_j$) in the mean (variance) model,
and write $\pi^\mu=(\pi^\mu_1,...,\pi^\mu_p)^T$, $\pi^\s=(\pi^\s_1,...,\pi^\s_q)^T$.
We assume that the inclusions of predictors are independent a priori with
\beqn
p(C|\pi^\mu)=\prod_{i\in C}\pi^\mu_i\prod_{i\not\in C}(1-\pi^\mu_i),\;\; p(V|\pi^\s)=\prod_{j\in V}\pi^\s_j\prod_{j\not\in V}(1-\pi^\s_j),
\eeqn
and the prior probability of a model $m$ with index sets $C$ and $V$ in its mean and variance models
is assumed to be
\beq\label{modelprior}
p(m)=p(C,V|\pi^\mu,\pi^\s)=p(C|\pi^\mu)p(V|\pi^\s).
\eeq
If no such detailed prior information is available for each individual predictor
(which is the situation we consider in this paper),
one may assume that $\pi^\mu_1=...=\pi^\mu_p=\pi_\mu$ and $\pi^\s_1=...=\pi^\s_q=\pi_\s$ 
(we note a slight abuse of notation here). Then
\beq\label{modelprior1}
p(C|\pi_\mu)=\pi_\mu^{|C|}(1-\pi_\mu)^{p-|C|},\;\; p(V|\pi_\s)=\pi_\s^{|V|}(1-\pi_\s)^{q-|V|},
\eeq
where hyperparameters $\pi_\mu,\ \pi_\s\in[0,1]$ are user-specified.
One can encourage parsimonious models by setting small ($<1/2$) $\pi_\mu$ and $\pi_\s$.
The smaller the $\pi_\mu$ and $\pi_\s$, the smaller prior probabilities are put on complex models.
By setting $\pi_\mu=\pi_\s=1/2$, one can set a uniform prior on the models. 
Another option is to put uniform distributions on $\pi_\mu$ and $\pi_\s$. Then
\beq\label{EBIC}
p(C)=\int_0^1 p(C|\pi_\mu)d\pi_\mu \propto\binom{p}{|C|}^{-1},\;\;p(V)=\int_0^1 p(V|\pi_\s)d\pi_\s \propto\binom{q}{|V|}^{-1}.
\eeq
This prior agrees with the one used in the extended BIC proposed by \cite{Chen:2008}.
It has the advantage of requiring no hyperparameter while still encouraging parsimony. 
We recommend using this as the default prior.

We now consider adding a single variable in either the mean or the
variance model, and then a one-step update to the current variational lower bound in the proposed model
as a computationally thrifty way of ranking the predictors for possible inclusion.
In our one-step update, we consider a variational
approximation in which the variational posterior distribution factorizes into separate parts for the added 
parameter and the parameters in the current model, as well as the factorization of mean and
variance parameters in Section 2.  We stress that this factorization is only assumed for the purpose of ranking
predictors for inclusion. Once a variable has been selected for inclusion, the posterior distribution
is approximated using the method outlined in Section 2.  
Write $\beta_C$ for the parameters in the current mean model and $X_C$ for the corresponding design matrix, 
and $\alpha_V$ for the parameters in the current variance model with $Z_V$ the corresponding design matrix.
Write $x_{Ci}$ for the $i$th row of $X_C$ and $z_{Vi}$ for the $i$th row of $Z_V$.  

\subsection{Ranking predictors in the mean model}
Let us consider first the effect of adding the predictor $x_j$, $j\in D\setminus C$, to the mean model.  
We write $\beta_j$ for the coefficient of $x_j$ and we consider a variational approximation to the 
posterior of the form
\beq\label{factorize}
q(\theta)=q(\beta_C)q(\beta_j)q(\alpha_V),
\eeq
with $q(\beta_C)=N(\mu_{\beta C}^q,\Sigma_{\beta C}^q)$, $q(\alpha_V)=N(\mu_{\alpha V}^q,\Sigma_{\alpha V}^q)$
and $q(\beta_j)=N(\mu_{\beta j}^q,(\sigma_{\beta j}^q)^2)$.  Suppose that we have fitted a variational
approximation for the current model (i.e. the model without $x_j$) using the procedure of Section 2.  
We now consider fitting the extended model with $\mu_{\beta C}^q,\Sigma_{\beta C}^q,\mu_{\alpha V}^q$ 
and $\Sigma_{\alpha V}^q$
fixed at the optimized values obtained for the current model, and consider just one step of a variational
algorithm for maximizing the variational lower bound in the new model with respect to the parameters
$\mu_{\beta j}^q, (\sigma_{\beta j}^q)^2$.  In effect for our variational lower bound (\ref{hetlowerbd}),
we are assuming that the variational posterior distribution for $({\beta_C}^T,\beta_j)^T$ is normal with 
mean $({\mu_{\beta C}^q}^T,\mu_{\beta j}^q)^T$
and covariance matrix 
\beqn
\left[ \begin{array}{cc}
 \Sigma_{\beta C}^q & 0 \\
 0 & (\sigma_{\beta j}^q)^2 
 \end{array}\right].
\eeqn
Substituting these forms into (\ref{hetlowerbd}) and further assuming $\mu_\beta^0=0$, $\mu_\alpha^0=0$, 
$\Sigma_\beta^0=\sigma_\beta^2 I$ and $\Sigma_\alpha^0=\sigma_\alpha^2 I$ (see the remarks at the end of
Section 2), we obtain the lower bound
\beq
L=L_{\text{old}}+\frac{1}{2}+\frac{1}{2}\log\frac{(\sigma_{\beta j}^q)^2}{\sigma_\beta^2}
-\frac{(\sigma_{\beta j}^q)^2}{2\sigma_\beta^2}-\frac{(\mu_{\beta j}^q)^2}{2\sigma_\beta^2}
-\frac{1}{2}\sum_{i=1}^n \frac{x_{ij}^2(\sigma_{\beta j}^q)^2+x_{ij}^2(\mu_{\beta j}^q)^2-
2x_{ij} \mu_{\beta j}^q (y_i-x_{Ci}^T\mu_{\beta C}^q)}
{\exp\left(z_{i V}^T\mu_{\alpha V}^q-\frac{1}{2}z_{i V}^T\Sigma_{\alpha V}^q z_{i V}\right)}, \label{hetlowerbd2}
\eeq
where $L_{\text{old}}$ is the previous lower bound for the current model without predictor $j$.  
Here we are writing $x_{ij}$ for the value of predictor $j$ for observation $i$.  Optimizing the above bound
with respect to $\mu_{\beta j}^q$ and $(\sigma_{\beta j}^q)^2$
and writing ${\hat{\mu}_{\beta j}}^q$ and $(\hat\sigma_{\beta j}^q)^2$
for the optimizers gives
\beq
\hat{\mu}_{\beta j}^q  = \left({\sum_{i=1}^n \frac{x_{ij}(y_i-x_{C i}^T\mu_{\beta C}^q)}
{\exp\left(z_{Vi}^T\mu_{\alpha V}^q-\frac{1}{2}z_{Vi}^T \Sigma_{\alpha V}^q z_{Vi}\right)}}\right)\Big/
\left({\frac{1}{\sigma_\beta^2}+\sum_{i=1}^n \frac{x_{ij}^2}
{\exp\left(z_{Vi}^T\mu_{\alpha V}^q-\frac{1}{2}z_{Vi}^T\Sigma_{\alpha V}^q z_{Vi}\right)}}\right)\label{optmub}
\eeq
and
\beq
({\hat\sigma}_{\beta j}^q)^2 =  \left(\frac{1}{\sigma_\beta^2}+\sum_{i=1}^n \frac{x_{ij}^2}
{\exp\left(z_{Vi}^T\mu_{\alpha V}^q-\frac{1}{2}z_{Vi}^T\Sigma_{\alpha V}^q z_{Vi}\right)}\right)^{-1}.\label{optsv}
\eeq
Substituting these back into the lower bound (\ref{hetlowerbd2}) gives
\beq
L_{\text{old}}+\frac{1}{2}\log\frac{({\hat\sigma}_{\beta j}^q)^2}{\sigma_\beta^2}
+\frac12\frac{({\hat\mu}_{\beta j}^q)^2}
{({\hat\sigma}_{\beta j}^q)^2}. \label{hetlowerbd3}
\eeq
If the variance model contains only an intercept, this result
agrees with greedy selection algorithms where predictors are ranked according to the correlation between
a predictor and the residuals from the current model (see, e.g. \cite{Zhang:2009}).  
We will discuss this point in detail in Section \ref{secHomo}.
Later we write the optimized value of
(\ref{hetlowerbd2}) as $L_j^M(C,V)$, the superscript $M$ means the lower bound associated with the model for {\em m}ean.   

\subsection{Ranking predictors in the variance model}\label{SecrankV}
So far we have considered only the addition of a predictor in the mean model.  We now attempt 
a similar analysis of the effect of inclusion of a predictor in the variance model.  With the mean
model fixed, suppose that we are considering adding a predictor $z_j$, $j\in E\setminus V$, to the variance model.  
We consider a normal approximation to the posterior
$q(\theta)=q(\beta_C)q(\alpha_V)q(\alpha_j)$ with $q(\beta_C)=N(\mu_{\beta C}^q,\Sigma_{\beta C}^q)$, 
$q(\alpha_V)=N(\mu_{\alpha V}^q,\Sigma_{\alpha V}^q)$ and 
$q(\alpha_j)=N(\mu_{\alpha j}^q,(\sigma_{\alpha j}^q)^2)$.
The variational lower bound is
\begin{align}
L_\text{old}&+\frac{1}{2}+\frac{1}{2}\log\frac{(\sigma_{\alpha j}^q)^2}{\sigma_{\alpha}^2} 
      -\frac{(\sigma_{\alpha j}^q)^2}{2\sigma_\alpha^2}-\frac{(\mu_{\alpha j}^q)^2}{2 \sigma_\alpha^2}
     -\frac{1}{2}\sum_i z_{ij}\mu_{\alpha j}^q  \nonumber \\
&-\frac{1}{2}\sum_{i=1}^n \left\{
    \frac{1}{\exp(z_{Vi}^T\mu_{\alpha V}^q-\frac{1}{2}z_{Vi}^T\Sigma_{\alpha V}^q z_{Vi}+z_{ij}\mu_{\alpha j}^q-\frac{1}{2}z_{ij}^2
    (\sigma_{\alpha j}^q)^2)}-\frac{1}{\exp(z_{Vi}^T\mu_{\alpha V}^q-\frac{1}{2}z_{Vi}^T\Sigma_{\alpha V}^q z_{Vi})}\right\}\times\nonumber\\
&\phantom{cccccccc}\left( (y_i-x_{Ci}^T\mu_{\beta C}^q)^2+x_{Ci}^T\Sigma_{\beta C}^q x_{Ci}\right),  \label{LD}
\end{align}
where $L_\text{old}$ is the lower bound for the current model without predictor $z_j$.  
To obtain good values for $\mu_{\alpha j}^q$ and $(\sigma_{\alpha j}^q)^2$,
we use an approximation similar to the one used for the variance parameters in Section 2.  
If we do not assume a normal form for $q(\alpha_j)$ but just the factorization 
$q(\theta)=q(\beta_C)q(\alpha_V)q(\alpha_j)$ and with the current $q(\beta_C)$ and 
$q(\alpha_V)$ fixed, then the optimal $q(\alpha_j)$ is
\beqn
q(\alpha_j)\propto \exp(E(\log p(\alpha_j)+\log p(y|\theta))),
\eeqn
where the expectation is with respect to $q(\beta_C)q(\alpha_V)$.  We have that
\bea
 E(\log p(\alpha_j)+\log p(y|\theta)) & = & 
 E\left(-\frac{1}{2}\log 2\pi-\frac{1}{2}\log \sigma_{\alpha}^2-\frac{\alpha_j^2}{2\sigma_\alpha^2}-\frac{n}{2}\log 2\pi-\frac{1}{2}\sum_{i=1}^n z_{Vi}^T\alpha_V \right. \nonumber \\
 &  & \left. -\frac{1}{2}\sum_{i=1}^n z_{ij}\alpha_j-\frac{1}{2}\sum_{i=1}^n \frac{(y_i-x_{Ci}^T\beta_C)^2}{\exp\left(z_{Vi}^T\alpha_V+z_{ij}\alpha_j \right)}\right) \nonumber  \\
  & = & -\frac{1}{2}\log 2\pi-\frac{1}{2}\log \sigma_{\alpha}^2-\frac{\alpha_j^2}{2\sigma_\alpha^2}-\frac{n}{2}\log 2\pi-\frac{1}{2}\sum_{i=1}^n z_{Vi}^T\mu_{\alpha V}^q \nonumber \\
  & &  -\frac{1}{2}\sum_{i=1}^n z_{ij}\alpha_j-\frac{1}{2}\sum_{i=1}^n \frac{(y_i-x_{Ci}^T\mu_{\beta C}^q)^2+x_{Ci}^T\Sigma_{\beta C}^q x_{Ci}}{\exp\left(z_{Vi}^T\mu_{\alpha V}^q+z_{ij}\alpha_j-\frac{1}{2}z_{Vi}^T\Sigma_{\alpha V}^q z_{Vi}\right)}. \label{logapproxn}
\eea
We make a normal approximation $N(\mbox{$\hat{\mu}_{\alpha j}^q$},(\hat{\sigma}_{\alpha j}^q)^2)$ 
to the optimal $q(\alpha_j)$ via the mode and negative inverse second derivative
of (\ref{logapproxn}).  
Differentiating with respect to $\alpha_j$, we obtain
\beqn
-\frac{\alpha_j}{\sigma_\alpha^2}-\frac{1}{2}\sum_{i=1}^n z_{ij}+\frac{1}{2}\sum_{i=1}^n \frac{z_{ij} v_i}
{\exp(z_{ij}\alpha_j)}
\;\;\text{where}\;\;
v_i=\frac{(y_i-x_{Ci}^T\mu_{\beta C}^q)^2+x_{Ci}^T\Sigma_{\beta C}^q x_{Ci}}{\exp\left(z_{Vi}^T\mu_{\alpha V}^q-\frac{1}{2}z_{Vi}^T\Sigma_{\alpha V}^q z_{Vi}\right)}.
\eeqn
Approximating $\exp(-z_{ij}\alpha_j)\approx 1-z_{ij}\alpha_j$ (i.e. using a Taylor
series expansion about zero), setting the derivative to zero and solving gives  
\beq\label{muqaj}
\mbox{$\hat{\mu}_{\alpha j}^q$}=\left(\frac{1}{2}\sum_{i=1}^n z_{ij}(v_i-1)\right)\Big/\left(\frac{1}{\sigma_\alpha^2}+\frac{1}{2}\sum_{i=1}^n z_{ij}^2 v_i\right).
\eeq
To get more accurate estimation of the mode,
some optimization procedure may be used here with \eqref{muqaj} used as an initial point.
In our R implementation, the Newton method was used because \eqref{logapproxn} has its second
derivative available in a closed form (see \eqref{sigmaqaj} below).
We found that \eqref{muqaj} is a very good approximation
as the Newton iteration very often stops after a small number of iterations
(with a stopping tolerance as small as $10^{-10}$).

Differentiating \eqref{logapproxn} once more, and finding the negative inverse
of the second derivative at $\mbox{$\hat{\mu}_{\alpha j}^q$}$ gives
\beq\label{sigmaqaj}
(\hat{\sigma}_{\alpha j}^q)^2=
\left(\frac{1}{\sigma_\alpha^2}+\frac{1}{2}\sum_{i=1}^n \frac{z_{ij}^2 v_i}
{\exp(z_{ij}\mbox{$\hat{\mu}_{\alpha j}^q$})}\right)^{-1}.
\eeq
We can plug these values back into the lower bound in order to rank different
predictors for inclusion in the variance model.  Once again we note the computationally
thrifty nature of the calculations.  We write the optimized value of (\ref{LD})
as $L_j^D(C,V)$, the superscript $D$ means the lower bound associated with the model for standard {\em d}eviance.  

\subsection{Summary of the algorithm}
We summarize our variable selection algorithm below.  We write $L(C,V)$ for the optimized value
of the lower bound (\ref{hetlowerbd}) with the predictor set $C$ in the mean model and the
predictor set $V$ in the variance model. 
Write $C_{+j}$ for the set $C\cup\{j\}$ and $V_{+j}$ for the set $V\cup\{j\}$.  

\paradot{Algorithm 2: Variational approximation ranking (VAR) algorithm}
\begin{enumerate}
\item Initialize $C$ and $V$ and set $L_\text{opt}:=L(C,V)$.
\item Repeat the following steps until stop
\begin{enumerate}
\item Store $C_\text{old}: = C$, $V_\text{old}: = V$. 
\item Let $j^*=\arg \max_j\{L_j^M(C,V)+\log p(C_{+j},V)\}$. 
If $L(C_{+j^*},V)+\log p(C_{+j^*},V)>L_\text{opt}+\log p(C,V)\}$ then set $C:=C_{+j^*}$, $L_\text{opt}=L(C_{+j^*},V)$.
\item Let $j^*=\arg \max_j \{L_j^D(C,V)+\log p(C,V_{+j})\}$. 
If $L(C,V_{+j^*})+\log p(C,V_{+j^*})>L_\text{opt}+\log p(C,V)$ then set $V:=V_{+j^*}$, $L_\text{opt}=L(C,V_{+j^*})$.
\item If $C=C_\text{old}$ and $V=V_\text{old}$ then stop, else return to (a).
\end{enumerate}
\end{enumerate}

\subsection{Forward-backward ranking algorithm}
The ranking algorithm described above can be regarded as a forward greedy algorithm
because it considers adding at each step another predictor to the current model.
Hereafter we refer to this algorithm as forward variational ranking algorithm or fVAR in short. 
Like the other forward greedy algorithms that have been widely used in many scientific fields,
the fVAR works well in most of the examples that we have encountered.
However, a major drawback with the forward selection algorithms is that 
if a predictor has been wrongly selected then it can not be removed anymore.
A natural remedy for this is to add a backward elimination process
in order to correct mistakes made in the earlier forward selection.
We present here a recipe for ranking predictors for exclusion in the mean and variance models.

Let $C,\ V$ be the current sets of predictors in the mean and variance models respectively.
With $j\in C$, we write $C_{-j}$ for the set $C\setminus\{j\}$ and consider now the effect
of removing the predictor $x_j$ to the lower bound.
In order to reduce computational burden,
we need some way to avoid the need to do lower bound maximization 
for each model $C_{-j}$ when ranking $x_j$ for exclusion. 
Similar as before, we consider a variational approximation using the factorization \eqref{factorize} for the  
variational posterior distribution.
Following steps \eqref{hetlowerbd2}-\eqref{hetlowerbd3}, 
we can approximately write the lower bound for the current model (i.e. the model contains $x_j$)
as the sum of the lower bound for the model without $x_j$ and a $x_j$-based term
\beq
L(C,V) \approx L(C_{-j},V)+\G^M_{C_{-j},V}(j) \label{lbfactorizeM},
\eeq
with
\beq
\G^M_{C_{-j},V}(j):=\frac{1}{2}\log\frac{({\hat\sigma}_{\beta j}^q)^2}{\sigma_\beta^2}
+\frac12\frac{({\hat\mu}_{\beta j}^q)^2}{({\hat\sigma}_{\beta j}^q)^2},
\eeq 
where ${\hat\mu}_{\beta j}^q,\ {\hat\sigma}_{\beta j}^q$ are as in \eqref{optmub} and \eqref{optsv} with $C$ replaced by $C_{-j}$.
All the relevant quantities needed in the calculation of $\G^M_{C_{-j},V}(j)$ are 
fixed at optimized values maximizing the lower bound for the current model.
The subscripts $C_{-j},V$ emphasize that 
the quantities needed are adjusted correspondingly 
when the predictor $j$ is removed from the mean model. 
The most plausible candidate for exclusion from the current mean model then is
\beq\label{candidate1}
j^* = \argmax_{j\in C}\{L(C_{-j},V)+\log p(C_{-j},V)\} = \argmin_{j\in C}\{\G^M_{C_{-j},V}(j)-\log p(C_{-j},V)\}. 
\eeq
We now rank the predictors for exclusion in the variance model.
Following the arguments in Section \ref{SecrankV} and the above,
we can write
\beq
L(C,V) \approx L(C,V_{-j})+\G^D_{C,V_{-j}}(j), \label{lbfactorizeD}
\eeq
with
\begin{align}
\G^D_{C,V}(j)&=\frac{1}{2}+\frac{1}{2}\log\frac{(\hat{\sigma}_{\alpha j}^q)^2}{\sigma_{\alpha}^2} 
      -\frac{(\hat{\sigma}_{\alpha j}^q)^2}{2\sigma_\alpha^2}-\frac{(\hat{\mu}_{\alpha j}^q)^2}{2 \sigma_\alpha^2}
     -\frac{1}{2}\sum_i z_{ij}\hat{\mu}_{\alpha j}^q  \nonumber \\
&-\frac{1}{2}\sum_{i=1}^n \left\{
    \frac{1}{\exp(z_{Vi}^T\mu_{\alpha V}^q-\frac{1}{2}z_{Vi}^T\Sigma_{\alpha V}^q z_{Vi}+z_{ij}\hat{\mu}_{\alpha j}^q-\frac{1}{2}z_{ij}^2
    (\hat{\sigma}_{\alpha j}^q)^2}-\frac{1}{\exp(z_{Vi}^T\mu_{\alpha V}^q-\frac{1}{2}z_{Vi}^T\Sigma_{\alpha V}^q z_{Vi})}\right\}\times\nonumber\\
&\phantom{cccccccc}\left( (y_i-x_{Ci}^T\mu_{\beta C}^q)^2+x_{Ci}^T\Sigma_{\beta C}^q x_{Ci}\right),
\end{align}
where $\hat{\mu}_{\alpha j}^q,\ \hat{\sigma}_{\alpha j}^q$ are as in \eqref{muqaj}-\eqref{sigmaqaj}
with $V$ replaced by $V_{-j}$.
The most plausible candidate for exclusion from the current variance model then is
\beq\label{candidate2}
j^* = \argmax_{j\in V}\{L(C,V_{-j})+\log p(C,V_{-j})\} = \argmin_{j\in V}\{\G^D_{C,V_{-j}}(j)-\log p(C,V_{-j})\}. 
\eeq
 
\paradot{Algorithm 3: Forward-backward variational approximation ranking algorithm}
\begin{enumerate}
\item Initialize $C$ and $V$, and set $L_\text{opt}=L(C,V)$.
\item Forward selection: as in Step 2 in Algorithm 2.
\item Backward elimination: Repeat the following steps until stop
\begin{enumerate}
\item Store $C_\text{old}: = C$, $V_\text{old}: = V$. 
\item Find $j^*$ as in \eqref{candidate1}. If $L(C_{-j^*},V)+\log p(C_{-j^*},V)>L_\text{opt}+\log p(C,V)$ then set $C=C_{-j^*}$, $L_\text{opt}=L(C_{-j^*},V)$.
\item Find $j^*$ as in \eqref{candidate2}. If $L(C,V_{-j^*})+\log p(C,V_{-j^*})>L_\text{opt}+\log p(C,V)$ then set $V=V_{-j^*}$, $L_\text{opt}=L(C,V_{-j^*})$.
\item If $C=C_\text{old}$ and $V=V_\text{old}$ then stop, else return to (a).
\end{enumerate}
\end{enumerate}
Hereafter we refer to this algorithm as fbVAR. 

In some applications where $X\equiv Z$, it might be meaningful 
to restrict the search for inclusion in the variance model
to those predictors that have been included in the mean model.
To this end, in the forward selection we just need to restrict 
the search for the most plausible candidate $j^*$ in Step 2(c) of Algorithm 2 to set $C$, 
i.e. $j^*=\arg \max_{j\in C} \{L_j^D(C,V)+\log p(C,V_j)\}$.
Also, when considering the removal of a candidate $j$ from the mean model in the backward elimination,
we need to remove $j$ from the variance model as well if $j\in V$,
i.e. Step 3(b) of Algorithm 3 must be modified to
\begin{itemize}
\item[3(b')]Let $j^*= \argmin_{j\in C}\{\G^M_{C_{-j},V_{-j}}(j)-\log p(C_{-j},V_{-j})\}$. If $L(C_{-j^*},V_{-j^*})+\log p(C_{-j^*},V_{-j^*})>L_\text{opt}+\log p(C,V)$ then set $C=C_{-j^*}$, $V=V_{-j^*}$, $L_\text{opt}=L(C_{-j^*},V_{-j^*})$.
\end{itemize}

Later we compare with the variable selection approaches for heteroscedastic regression
implemented in the GAMLSS (generalized additive model for location, scale and shape) package \citep{Rigby:2005}.
The GAMLSS framework allows modeling of the mean and other parameters (like the standard deviation, skewness
and kurtosis) of the response distribution as flexible functions of predictors.
Variable selection is done with stepwise selection using a generalized AIC or BIC as the stopping rule.
The GAMLSS uses a Fisher scoring algorithm to maximize the likelihood
for ranking every predictor for inclusion/exclusion rather than only the most plausible one 
as in the VAR algorithm, which leads to a heavy computational burden for large-$p$ problems.

\subsection{The ranking algorithm for homoscedastic regression}\label{secHomo}
In order to get more insight into our VAR algorithm,
we discuss in this section the algorithm for the homoscedastic linear regression model.
In the case of constant variance, the variance parameter $\a$ now becomes scalar,
we rename the quantities $\Sigma_\a^0,\ \Sigma_\a^q$ as $(\sigma_\a^0)^2,\ (\sigma_\a^q)^2$ respectively.
The optimal choice \eqref{optdis1} for $p(\a)$ becomes
\beqn
q(\a)\propto \exp\left(-\frac{n}{2}\a-\frac12ve^{-\a}-\frac12\frac{\a^2}{(\sigma_\a^0)^2}\right)\;\;\text{where}\;\;v:=\sum_{i=1}^n\left((y_i-x_i^T\mu_\b^q)^2+x_i^T\Sigma_\b^qx_i\right).
\eeqn
Using the approximation $\exp({-\a})\approx 1-\a$, it is easy to see that the mean and variance of the normal approximation are
\beqn
\mu_\a^q=\frac{v-n}{v+2/(\sigma_\a^0)^2}\;\;\text{and}\;\;(\sigma_\a^q)^2=\left(\frac{v}{2}e^{-\mu_\a^q}+\frac{1}{(\sigma_\a^0)^2}\right)^{-1}
\eeqn
respectively. We now can replace Steps 4 and 5 in Algorithm 1 by these two closed forms 
so that the computations can be reduced greatly.
Similar to the discussion in Section \ref{SecrankV},
the Newton method may be used here in order to get a more accurate estimate of the mode.
In our experience, however, this is not necessary here.  

For the variable selection problem, we now just need to rank the predictors for inclusion/exclusion in the mean model.
Assume that we are using the uniform model prior, i.e. $p(C,V)\equiv$constant,
or a model prior as in \eqref{modelprior1},
the ranking of predictors then follows the ranking of lower bounds.
We further assume that the design matrix $X$ has been standardized such that $\sum_ix_{ij}^2=n$,
the optimizer $(\hat\sigma_{\b j}^q)^2$ in \eqref{optsv} does not depend on $j$,
and the ranking of the lower bound \eqref{hetlowerbd3}
follows the ranking of $\left| \sum_{i=1}^n x_{ij}(y_i-x_{Ci}^T\mu_{\beta C}^q)\right|$
(i.e. it follows the ranking of the absolute correlation of the predictors with the standardized
residuals from the current model).   
This result agrees with frequentist matching pursuit and greedy algorithms 
where predictors are ranked according to the correlation between
a predictor and the residuals from the current model
\citep{Mallat:1993,Zhang:2009,Efron:2004}.  
This is also similar to computationally thrifty
path following algorithms \citep{Efron:2004,Zhao:2007}.

For the existing frequentist algorithms for variable selection,
extra tuning parameters, such as the shrinkage parameter in penalization procedures,
the number of iterations in matching pursuit and the stopping parameter $\epsilon$ in greedy algorithms,
need to be chosen. And their performance depends critically on the method used to choose these tuning parameters.
Our method does not require any extra tuning parameters.
The final model is chosen when the lower bound (plus the log model prior) is maximized
 - a widely used stopping rule in Bayesian model selection with variational Bayes 
\citep{Beal:2003,Beal:2006,Wu:2010}.

Unlike many commonly used greedy algorithms, 
our Bayesian framework is able to incorporate prior information (if available) on models
and/or to encourage parsimonious models if desired. 
Besides involving extra tuning parameters, penalized estimates are often biased
\citep{Friedman:2008,Efron:2004}.
While our method can penalize non-zero coefficients through the prior if desired,
it does not rely on shrinkage
of coefficients to do variable selection, so that in principle it might 
produce better estimation of non-zero coefficients.
Simulation studies in Section \ref{secSimul} confirm this point.  
Note that we do not consider models of all sizes, the algorithm stops when important predictors
have been included in the model, so that computations in Algorithm 1 
just involve matrices with low-dimension.
This is another advantage which makes our method potentially valuable 
for variable selection in high-dimensional problems.
Our experience shows that the VAR algorithm is as fast as the LARS algorithm
in problems with thousands of predictors.  

\section{Applications}\label{secApp}
\paradot{Example 1: biscuit dough data}
The biscuit dough data is a benchmark ``large $p$, small $n$" data set 
that was originally designed and analyzed in \cite{Osborne:1984}.
The purpose of this study was to investigate
the practical benefit of using near-infrared (NIR) spectroscopy
in the food processing industries.
In their experiment, the aim was to predict biscuit dough constituents 
based on near-infrared spectrum of dough samples.
The four constituents of interest were fat, sucrose, flour and water.
Two data sets (training set $D^T$ and prediction or validation set $D^P$) 
were made up in the same manner
in which the percentages of four constituents were exactly calculated. 
These percentages serve as response variables.
There were 39 samples in the training set and 31 in the prediction set.
The NIR spectrum of dough pieces was measured at equally spaced wavelengths
from 1100 to 2498 nanometers (nm) in steps of 2 nm.
Following \cite{Brown:2001}, we removed the first 140 and last 49 wavelengths
because they were thought to contain little useful information.
From remaining wavelengths ranging from 1380 to 2400 nm, every second wavelength was considered, 
which increases the space step to 4 nm.
The final data sets consist of 256 predictors and four responses
which were treated separately in four univariate linear regression models
rather than in a single multivariate model.

\begin{figure*}
\centerline{\includegraphics[width=1\textwidth]{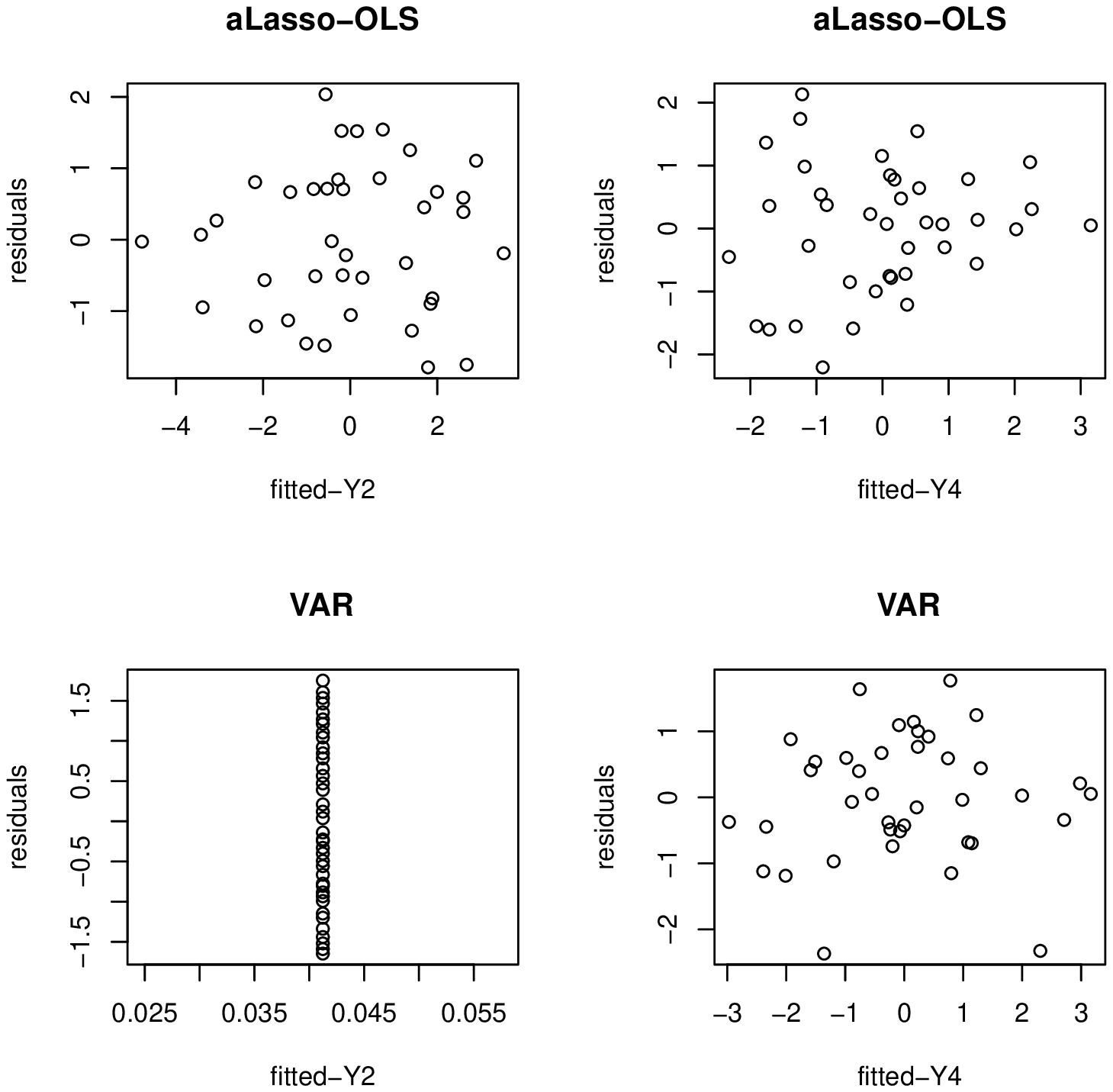}}
\caption{\label{biscuitfig1}
The biscuit dough data.}
\end{figure*}

The most popular and easiest way to check heteroscedasticity 
is to use plotting techniques.
When the OLS fit is valid,
plotting studentized residuals against fitted values is a powerful technique to use \citep{Carroll:1988}.
In our current case of ``large $p$, small $n$", we first used the adaptive Lasso (aLasso) 
of \cite{Zou:2006} to select likely predictors 
and then applied the above technique to the selected predictors.  
We name this method aLasso-OLS.
Figure \ref{biscuitfig1} shows the plots of aLasso-OLS studentized residuals 
versus fitted values (where homoscedasticity was assumed), 
and also the corresponding plots resulting from our fbVAR algorithm\footnote{We did not 
apply the restriction here, because there was no good reasons
to restrict the search for inclusion in the variance model to the predictors 
in the mean model. The search combined both forward and backward moves and the uniform model prior 
(i.e. $\pi_\mu=\pi_\sigma=1/2$) was used.}
(where heteroscedasticity was assumed)
for the response variables sucrose ($Y_2$) and water ($Y_4$);
all were calculated based on the training set.
The plots for the other responses were not shown because the need for a heteroscedastic model
was not visually obvious.
We can see that in general fitting the homoscedastic regression model to these responses was not appropriate here.
Looking at the aLasso-OLS plot for $Y_4$, for example,
there was clear evidence that (absolute values of) residuals decrease when fitted values increase,
and the heteroscedastic model estimated by the VAR method gave a more satisfying residual plot.
For the response $Y_2$, the VAR did not select any predictor (except the intercept) 
for inclusion in the mean model, although several predictors were selected for the variance model.
This ``non-null" variance model reflects the heteroscedasticity 
which is visualizable in the aLasso-OLS plot for $Y_2$.
The null model for the mean model was quite a surprise, 
since all the works analyzing $Y_2$ assuming the homoscedastic linear model 
that we are aware of in the literature reported non-null models.
The aLasso in our analysis selected only one predictor, the 130th.
Among the plots of all 4 responses against all selected predictors,
the plot of $Y_2$ against the selected predictor (by the aLasso of course) looked very random
compared to the others.
This in some sense supported visually the null mean model for $Y_2$. 
   
We then employed the resulting models to make predictions
and used the validation set $D^P$ to examine the appropriateness 
of assuming heteroscedasticity for this biscuit dough data.
The usefulness of a model was measured by two criteria:
one was the mean squared error of prediction defined as
\beqn
\text{MSE} = \frac{1}{|D^P|}\sum_{(x,y)\in D^P}\|y-\hat y(x)\|^2
\eeqn
and the other was the partial prediction score
\beqn
\text{PPS} = \frac{1}{|D^P|}\sum_{(x,y)\in D^P}-\log \hat p(y|x),
\eeqn
where $\hat p(.)$ is the density estimated under the model.
It is understood that the smaller the MSE and PPS, the better the model.
The MSE and PPS evaluated on the 31 samples of the validation set for the aLasso and VAR methods
are summarized in Table \ref{biscuittable}.
Except for the case of $Y_2$ (sucrose), the heteroscedastic models estimated by the VAR method
had a big improvement over the homoscedastic models estimated by the aLasso.
The poor predictive performance of the VAR (and the aLasso as well) on $Y_2$
was probably due to the reasons discussed above: 
there was no linear relationship between the NIR spectrum and the sucrose constituent.

\begin{table}
  \begin{center}
    \begin{tabular}{c|c|c|c|c|}
&\multicolumn{2}{c|}{MSE}&\multicolumn{2}{c|}{PPS}\\
\cline{2-5}
&aLasso&VAR&aLasso&VAR\\
\hline	
fat	&2.61	&0.09	&1.91	&0.25\\
sucrose	&13.56	&14.87	&2.73	&2.77\\
flour	&4.43	&0.79	&2.16	&1.37\\
water	&0.64	&0.18	&1.20	&0.64
\end{tabular}
\end{center}
  \caption{The biscuit dough data: MSE and PPS evaluated on the validation set for the aLasso and VAR methods.}\label{biscuittable}
\end{table}

This biscuit dough data was also carefully analyzed in \cite{Brown:2001} using
a Bayesian wavelet regression framework. They first used a wavelet transformation
to transform the original predictors to wavelet coefficients
and then applied a Bayesian (homoscedastic) regression approach 
to regress the responses on the derived wavelet coefficients.
Prediction was done using Bayesian model averaging (BMA) over a set of 500 likely models,
and MSE values were reported to be 0.06, 0.45, 0.35 and 0.05 respectively.
This methodology seems not comparable to ours because
(i) it was conducted based on wavelet coefficients rather than the original predictors and
(ii) prediction was done using BMA rather than a single selected model.

Because the four response variables are percentages and sum to 100,
an anonymous reviewer raised a concern about spurious correlations between them.
While this may be a concern for a multivariate analysis,
we treated the four responses separately in four univariate linear regression models
rather than in a single multivariate model,
so that compositional effects would not be a problem here.
To justify this, we considered the following transformation \citep{Aitchison:1986} of the responses
\beqn
U_i=\log\frac{Y_i}{Y_3},\;\;i=1,2,4.
\eeqn   
The choice of $Y_3$ for denominator was natural because the flour is a major constituent \citep{Brown:2001}.
After fitting the regression models with these three new responses,
the fitted and predicted values were transformed back to the original scale via
\beqn
Y_i=\frac{100\exp(U_i)}{\sum\exp(U_i)+1},\;\;i=1,2,4\;\;\text{and}\;\;Y_3=\frac{100}{\sum\exp(U_i)+1}.
\eeqn
The MSE evaluated on the validation set for the aLasso were 4.71, 13.43, 7.07, 1.45
and for the VAR method were 2.16, 5.22, 3.36, 0.57.
We did not report PPS because it is not clear how to properly calculate PPS 
in the case of heteroscedastic regression for such tranformed data.
Comparing to the result in Table \ref{biscuittable},
it seems that the above transformation which is to account for potential compositional effects 
does not give a positive impact overall.
This result also agrees with the analysis of \cite{Brown:2001}.

\paradot{Example 2: diabetes data}
In the second application we applied the VAR method to analyzing a 
benchmark data set in the literature on progression of diabetes \citep{Efron:2004}.
Ten baseline variables, age, sex, body mass index, average blood pressure
and six blood serum measurements, were obtained for each of $n = 442$ diabetes
patients, as well as the response of interest $y$, a quantitative measure of disease
progression one year after baseline. 
We constructed a (heteroscedastic, if necessary) linear regression model 
to predict $y$ from these ten input variables.
In the hope of improving prediction accuracy, we considered a ``quadratic model" with 64 predictors.
We distinguish between input variables and predictors, 
for example, in a quadratic regression model on two input variables age and income, there
are five predictors (age, income, $\text{age}\times\text{age}$, $\text{income}\times\text{income}$ and $\text{age}\times\text{income}$).

The analysis of the full data set showed clear evidence of heteroscedasticity.
See again Figure \ref{solutionpath} for the solution paths resulting from our VAR algorithm with the uniform model prior
(where only forward selection was implemented and the search for inclusion in the variance model was restricted).
The VAR and GAMLSS both selected some predictors to include in the variance model.
Furthermore, there was quite a clear pattern in the plot 
of the OLS studentized residuals indicating heteroscedasticity (results not shown).
Interestingly, when fitting $y$ with only ten input variables as the predictors,
diagnostics and the selected model by VAR both showed no evidence of heteroscedasticity.
This result agreed with the homoscedasticity assumption often used in the literature 
for this diabetes data set.  

To assess predictive performance, we randomly selected 300 instances to form the training set, with
the remainder serving as the validation set.
Of 64 predictors, the VAR selected 13 to include in the mean model and 12 to include in the variance model,
while the GAMLSS selected 23 and 7 respectively.
Under the assumption of constant variance, the aLasso selected 43 predictors.  
On the validation set, the models estimated by aLasso, GAMLSS and VAR had PPS of 5.50, 15.93, 5.41
and MSE of 3264.95, 3506.32, 2993.16 respectively.
In order to reduce the uncertainty in training-validation separation,
we recorded the MSE and PPS over 50 such random partitions,
and obtained the averaged MSE for aLasso, GAMLSS and VAR of 3715.08 (641.56), 4069.81 (1681.70), 3082.78 (774.85)
and the averaged PPS of 5.84 (0.15), 56.72 (11.52), 5.76 (0.16) respectively.
The numbers in brackets are standard deviations.
The GAMLSS method performed poorly in this example but it should be stressed that we have
only used the default implementation (i.e. stepwise selection with both forward and backward moves
and the generalized AIC used as the stopping rule)
in the GAMLSS R package. 
Further experimentation with tuning parameters
in the information criterion might produce better results.

\section{Experimental studies}\label{secSimul}
In this section, we present experimental studies for our method.
We first compare the accuracy of our variational approximation algorithm 
to that of MCMC in approximating a posterior distribution.
We then compare the VAR method for variable selection to the aLasso and GAMLSS 
in both heteroscedastic and homoscedastic regression. 
In the examples described below, the EBIC prior \eqref{EBIC}
was used as a default prior.
This prior has very little impact in low-dimensional cases 
but considerable impact in high-dimensional cases
in terms of encouraging parsimony \citep{Chen:2008}.

\begin{figure*}
\centerline{\includegraphics[width=1\textwidth]{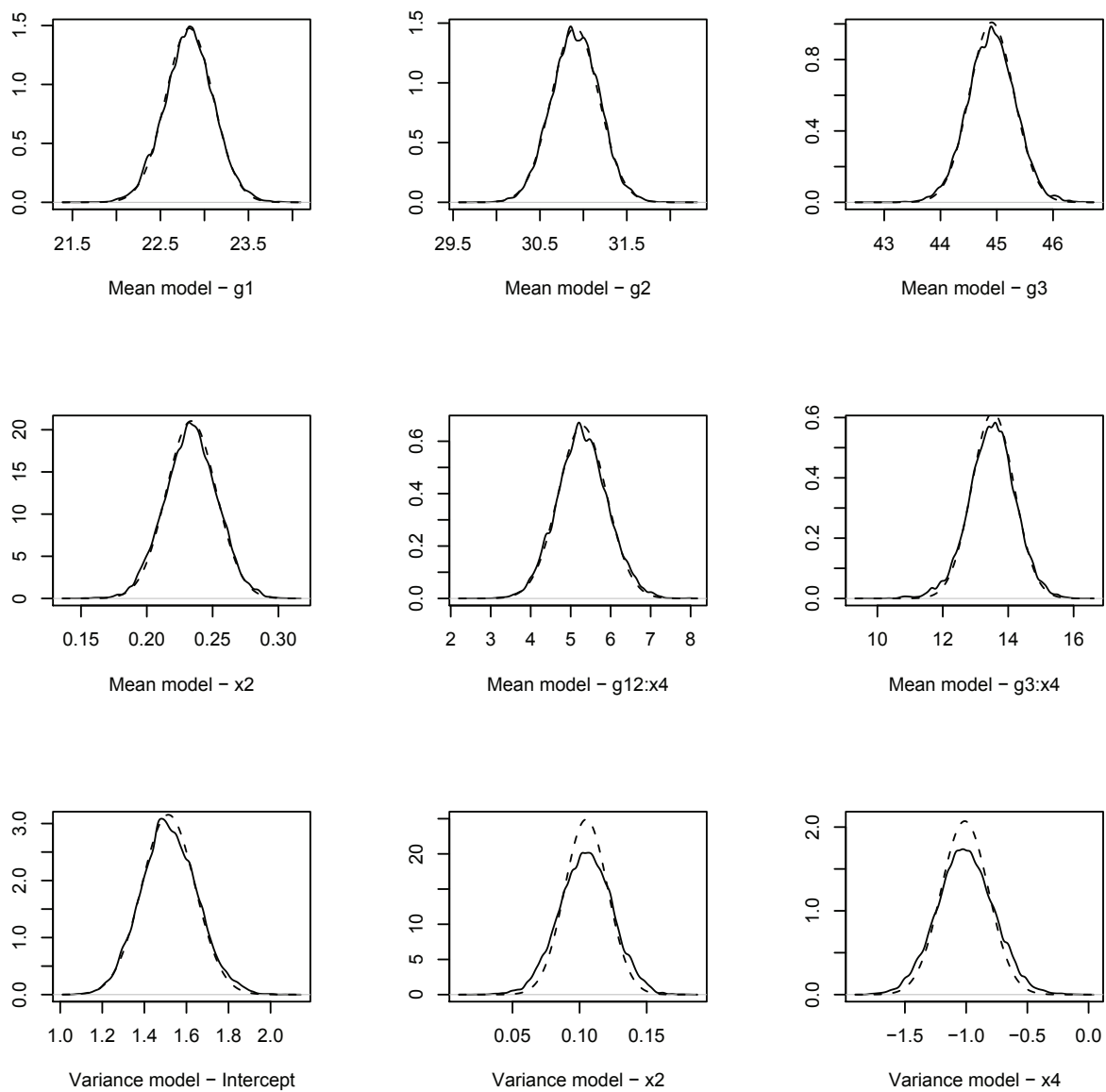}}
\caption{\label{sniffer} Estimated marginal posterior densities for coefficients in the mean and variance
models for the sniffer data. Solid lines are kernel estimates from MCMC samples from the
posterior and dashed lines are variational approximate marginal posterior densities.}
\end{figure*}

\paradot{The accuracy of the variational approximation}
In this example we demonstrate the accuracy of the variational approximation for describing
the posterior distribution in a heteroscedastic model, without considering the model selection aspects.  
We considered a data set described in \cite{Weisberg:2005}, see also \cite{Smyth:1989}.
The data were concerned with the hydrocarbon vapours which escape when petrol is pumped
into a tank. Petrol pumps are fitted with vapour recovery systems, which may not be completely
effective and ``sniffer" devices are able to detect if some vapour is escaping. An
experiment was conducted to estimate the efficiency of vapour recovery systems in which
the amount of hydrocarbon vapour given off, in grams, was measured, along with four predictor
variables. The four predictor variables were initial tank temperature ($x_1$), in degrees
Fahrenheit, the temperature of the dispensed gasoline ($x_2$), in degrees Fahrenheit, the initial
vapour pressure in the tank ($x_3$), in pounds per square inch, and the initial vapour pressure
of the dispensed gasoline ($x_4$), in pounds per square inch. \cite{Smyth:1989} considers fitting a
heteroscedastic linear model with the mean model
\beqn
\mu= \beta_1 g_1 + \beta_2 g_2 + \beta_3 g_3 + \beta_4 x_2 + \beta_5 g_{12}x_4 + \beta_6 g_3 x_4
\eeqn
and the variance model
\beqn
\log \sigma^2 = \alpha_0 + \alpha_1 x_2 + \alpha_2 x_4,
\eeqn
where $g_1$, $g_2$ and $g_3$ are three binary indicator variables for different ranges of $x_1$ and
$g_{12}=g_1+g_2$. In fitting the mean model the last three terms are orthogonalized with
respect to the first three, so that the coefficients of the indicators are simply group means
for the corresponding subsets of $x_1$, and in the variance model $x_2$ and $x_4$ were mean centered.
We considered our variational approximation to the posterior distribution in a Bayesian analysis
where the priors were multivariate normal with mean zero and covariance 10000$I$ for both
$\beta$ and $\alpha$.  Figure \ref{sniffer} shows estimated marginal posterior densities for all parameters in the
mean and variance models. The top two rows show the mean parameters and the bottom row
the variance parameters. The solid lines are kernel density estimates of the marginal posteriors
constructed from MCMC samples and the dotted lines are the variational approximate
posterior marginals. The mean and variance from the variational approximation were used
to define a multivariate Cauchy independence proposal for a Metropolis-Hastings scheme to
generate the MCMC samples. 100,000 iterations were drawn, with 1,000 discarded as ``burn
in". One can see that for the mean parameters, the variational approximation is nearly exact.
For the variance parameters, point estimation is very good, but there is a slight tendency
for the variational approximation to underestimate posterior variances. 
The final lower bound is -326.68, with agreement to two decimal places within the first two iterations 
and convergence after 5 iterations.  Compared to -326.5, the marginal likelihood computed
using the MCMC method of \cite{Chib:2001}, this lower bound is very tight.

\paradot{Heteroscedastic case} 
We present here a simulation study for our VAR method for simultaneous variable selection and parameter estimation
in heteroscedastic linear regression models,
and compare its performance to that of the GAMLSS and aLasso methods.
Data sets were generated from the model
\beq\label{linearmodel}
y = 2+x^T\tilde\b+\sigma e^{\frac12x^T\tilde\a}\epsilon,
\eeq
with $\tilde\b=(3,\ 1.5,\ 0,\ 0,\ 2,\ 0,\ 0,\ 0)^T$, $\epsilon\sim N(0,\ 1)$.
Predictors $x$ were first generated from normal distributions $N(0,\Sigma)$
with $\Sigma_{ij}=0.5^{|i-j|}$ and then transformed into the unit interval
by the cumulative distribution function $\Phi(.)$ of the standard normal.
The reason for making the transformation was to control the magnitude of noise level, 
i.e. the quantity $\sigma e^{\frac12x^T\tilde\a}$.  
Let $\b=(2,\ \tilde\b^T)^T$ and $\a=(\log\sigma^2,\tilde\a^T)^T$ be the mean and variance parameters respectively,
where $\tilde\a=(0,\ 3,\ 0,\ 0,\ -3,\ 0,\ 0,\ 0)^T$.
Note that the true predictors in the variance model were among those in the mean model.
This {\em prior} information was employed in the GAMLSS and VAR.

The performance was measured by correctly-fitted rates (CFR), numbers of zero-estimated coefficients (NZC) 
(for both mean and variance models), mean squared error (MSE) of predictions 
and partial prediction score (PPS) averaged over 100 replications.
MSE and PPS were evaluated based on independent prediction sets generated in the same manner
as the training set. 
We compared the performance of the VAR and GAMLSS methods (when heteroscedasticity was assumed) 
to that of the aLasso (when homoscedasticity was assumed).
The simulation results are summarized in Table \ref{hetero} for 
various factors sample size $n$, 
$n_P$ (size of prediction sets $D^P$) and $\sigma$.
As shown, the VAR method did a good job and outperformed the others.
 
We also considered a ``large $p$, small $n$" case in which $\tilde\b$ and $\tilde\a$
in model \eqref{linearmodel} were vectors of dimension 500 with most of the components zero
except $\tilde\b_{50}=\tilde\b_{100}=...=\tilde\b_{250}=5$, $\tilde\b_{300}=\tilde\b_{350}=...=\tilde\b_{500}=-5$
and $\tilde\a_{100}=\tilde\a_{200}=5$, $\tilde\a_{300}=\tilde\a_{400}=-5$.
The simulation results are summarized in Table \ref{hhetero}.
Note that the GAMLSS is not applicable when $n<p$,
and moreover that in the case with $n\geq p$ and with large $p$ the current implementation version of the GAMLSS 
is much more time consuming compared to the VAR
and even not working with $p$ as large as $500$,
since the package was not designed for such applications. 
We are not aware of any existing methods in the literature for variable selection
in heteroscedastic linear models for ``large $p$, small $n$" case.

\begin{table}
  \begin{center}
    \begin{tabular}{c|c|l|r|r|r}
$n=n_P$	&$\sigma$	&measures	&aLasso		&GAMLSS		&VAR\\
\hline\hline
50	&0.5		&CFR in mean	&64 (4.56)	&36 (4.06)	&80 (4.88)\\
	&		&CFR in var.	&nil		&70 (5.74)	&80 (5.96)\\
	&		&MSE		&0.56		&0.49		&0.48\\
	&		&PPS		&1.17		&0.89		&0.87\\
\cline{2-6}
	&1		&CFR in mean	&22 (4.72)	&38 (4.60)	&56 (5.00)\\
	&		&CFR in var.	&nil		&50 (5.88)	&60 (6.22)\\
	&		&MSE		&2.45		&2.29		&2.24\\
	&		&PPS		&2.01		&1.78		&1.69\\
\hline
100	&0.5		&CFR in mean	&74 (4.50)	&30 (3.98)	&88 (4.84)\\
	&		&CFR in var.	&nil		&64 (5.62)	&90 (5.90)\\
	&		&MSE		&0.52		&0.48		&0.48\\
	&		&PPS		&1.12		&0.87		&0.77\\
\cline{2-6}
	&1		&CFR in mean	&36 (4.68)	&42 (4.30)	&66 (4.76)\\
	&		&CFR in var.	&nil		&58 (5.72)	&76 (5.84)\\
	&		&MSE		&2.20		&2.08		&2.03\\
	&		&PPS		&1.83		&1.62		&1.51\\
\hline
200	&0.5		&CFR in mean	&94 (4.90)	&48 (4.14)	&100 (5.00)\\
	&		&CFR in var.	&nil		&70 (5.70)	&94 (5.94)\\
	&		&MSE		&0.48		&0.46		&0.46\\
	&		&PPS		&1.06		&0.87		&0.74\\
\cline{2-6}
	&1		&CFR in mean	&56 (4.36)	&36 (4.06)	&88 (4.88)\\
	&		&CFR in var.	&nil		&82 (5.80)	&100 (6.00)\\
	&		&MSE		&2.01		&1.93		&1.92\\
	&		&PPS		&1.77		&1.52		&1.42\\
\end{tabular}
\end{center}
  \caption{Small-$p$ case: CFR, NZC, MSE and PPS averaged over 100 replications. The numbers in parentheses are NZC.
The true number of non-zero coefficients in the mean model was 5 and in the variance model was 6.}\label{hetero}
\end{table}

\begin{table}
  \begin{center}
    \begin{tabular}{c|c|c|c|c|c|c|c|c|c}
&&\multicolumn{4}{c|}{VAR}&\multicolumn{3}{c|}{aLasso}\\
\cline{3-9}
$n=n_P$	&$\sigma$	&CFR in mean	&CFR in var.	&MSE	&PPS	&CFR in mean	&MSE	&PPS\\
\hline	
100	&0.5		&80 (489.75)	&90 (495.90)	&5.40	&1.91	&20 (491.80)	&11.65	&2.66\\
	&1		&70 (489.05)	&65 (495.80)	&20.29	&2.31	&0 (495.75)	&35.11	&3.28\\ 	
150	&0.5		&100 (490.00)	&95 (495.90)	&13.77	&0.85	&40 (491.95)	&20.02	&3.41\\
	&1		&95 (489.95)	&85 (495.85)	&28.97	&1.52	&5 (495.05)	&43.19	&3.69
\end{tabular}
\end{center}
  \caption{Large-$p$ case: CFR, NZC, MSE and PPS averaged over 100 replications. The numbers in parentheses are NZC.
The true number of non-zero coefficients in the mean model was 490 and in the variance model was 496.}\label{hhetero}
\end{table}

\paradot{Homoscedastic case} 
We also considered a simulation study 
when the data come from homoscedastic models.
Data sets were generated from the linear model \eqref{linearmodel} with $\tilde\a\equiv0$, i.e.
\beqn
y = 2+x^T\tilde\b+\sigma\epsilon
\eeqn
with predictors $x$ generated from normal distributions $N(0,\Sigma)$ with $\Sigma_{ij}=0.5^{|i-j|}$.
We were concerned with simulating a sparse, high-dimensional case.
To this end, $\tilde\beta$ was set to be a vector of 1000 dimensions with
the first 5 entries were $5,\ -4,\ 3,\ -2,\ 2$ and the rest were zeros.  
We used the modified ranking algorithm discussed in Section \ref{secHomo} 
with both forward and backward moves and the default prior \eqref{EBIC}.
The performance was measured as before by CFR, NZC and MSE
but MSE was defined as the squared error between the true vector $\b$ and its estimate. 
The simulation results are summarized in Table \ref{tablehomo}.
The big improvement of the VAR over the aLasso in this example is surprising and probably due to the
reasons discussed in Section \ref{secHomo}.

\begin{table}
  \begin{center}
    \begin{tabular}{c|c|c|c|c|c}
&&\multicolumn{2}{c|}{CFR (NZC)}&\multicolumn{2}{c}{MSE}\\
\cline{3-6}
$n=n_P$&$\sigma$&aLasso&VAR&aLasso&VAR\\
\hline
50	&	1	&0 (994.42)  &38 (994.34) &31.21 &17.72\\
	&	2	&0 (994.54)  &2  (992.36) &38.21 &33.16\\
\hline
100	&	1	&46 (995.62) &96 (994.96) &8.40 &0.09\\
	&	2	&16 (996.14) &32 (993.56) &11.87 &2.09\\
\hline
200	&	1	&90 (995.10) &98 (994.98) &6.34 &0.04\\
	&	2	&44 (995.56) &32 (993.40) &7.78 &0.62
\end{tabular}
\end{center}
  \caption{Homoscedastic case: CFR, MSE and NZC averaged over 100 replications for aLasso and VAR.
The true number of non-zero coefficients was 995.}\label{tablehomo}
\end{table}

\paradot{Remarks on calculations}
The VAR algorithm was implemented using R and the code is freely available on the authors' websites.
The weights used in the aLasso were assigned as usual as $1/|\hat\b_j|$ with $\hat\b_j$ being 
the MLE (when $p<n$) or the Lasso estimate (when $p\geq n$) of $\b_j$.
The tuning parameter $\lambda$ was selected by 5-fold cross-validation. 
The implementation of the aLasso and GAMLSS was carried out with the help of the R packages glmnet and gamlss.

\section{Concluding remarks}\label{discussion}
We have presented in this paper a strategy for variational lower bound maximization
in heteroscedastic linear regression,
and a novel fast greedy algorithm for Bayesian variable selection.
In the homoscedastic case with the uniform model prior, 
the algorithm reduces to widely used matching pursuit algorithms.
The suggested methodology has proven efficient, especially for high-dimensional problems.

Benefiting from the variational approximation approach - a fast deterministic alternative
and complement to MCMC methods for computation in high-dimensional problems - our methodology has potential for 
Bayesian variable selection in more complex frameworks.
A potential research direction is to extend the method
to simultaneous variable selection and number of experts selection in flexible regression
density estimation with mixtures of experts \citep{Geweke:2007,Villani:2009}. 
This research direction is currently in progress.
Another potential research direction is to extend the method to grouped variable selection. 

\section*{Appendix A}
Below we write $E_q(\cdot)$ for an
expectation with respect to the variational posterior.  In the notation of Section 1 we have
\bqan  
 T_1 & = & -\frac{p+q}{2}\log 2\pi -\frac{1}{2}\log |\Sigma_{\beta}^0|-\frac{1}{2}\log|\Sigma_{\alpha}^0| \\
  &  & -\frac{1}{2}E_q((\beta-\mu_{\beta}^0)^T { \Sigma_{\beta}^0}^{-1} (\beta -\mu_{\beta}^0))
       -\frac{1}{2}E_q((\alpha-\mu_{\alpha}^0)^T {\Sigma_{\alpha}^0}^{-1} (\alpha-\mu_{\alpha}^0)) \\
  & = & -\frac{(p+q)}{2}\log 2\pi -\frac{1}{2}\log |\Sigma_{\beta}^0|-\frac{1}{2}\log|\Sigma_{\alpha}^0| \\
  &   & -\frac{1}{2}\tr ({\Sigma_{\beta}^0}^{-1}\Sigma_{\beta}^q)
        -\frac{1}{2}\tr ({\Sigma_{\alpha}^0}^{-1}\Sigma_{\alpha}^q)
        -\frac{1}{2}(\mu_{\beta}^q-\mu_{\beta}^0)^T{\Sigma_{\beta}^0}^{-1}(\mu_{\beta}^q-\mu_{\beta}^0)  \\
  &   & -\frac{1}{2}(\mu_{\alpha}^q-\mu_{\alpha}^0)^T{\Sigma_{\alpha}^0}^{-1}(\mu_{\alpha}^q-\mu_{\alpha}^0),\\
T_2 & = & -\frac{n}{2}\log 2\pi-\frac{1}{2}E_q(\sum_{i=1}^n z_i^T\alpha)-\frac{1}{2}E_q\left(\sum_{i=1}^n\frac{(y_i-x_i^T\beta)^2}{\exp(z_i^T \alpha)}\right) \\
  & = & -\frac{n}{2}\log 2\pi-\frac{1}{2}\sum_{i=1}^n z_i^T \mu_\alpha^q-\frac{1}{2}\sum_{i=1}^n \frac{x_i^T\Sigma_{\beta}^q x_i+(y_i-x_i^T\mu_\beta^q)^2}{\exp\left(z_i^T\mu_\alpha^q-\frac{1}{2}z_i^T\Sigma_{\alpha}^q z_i\right)}
\eqan
and
\bqan
 T_3 & = & \frac{p+q}{2}\log 2\pi+\frac{1}{2}\log |\Sigma_{\beta}^q|+\frac{1}{2}\log |\Sigma_{\alpha}^q| \\
  & & +\frac{1}{2}E_q((\beta-\mu_\beta^q)^T{\Sigma_{\beta}^q}^{-1}(\beta-\mu_\beta^q))
      +\frac{1}{2}E_q((\alpha-\mu_\alpha^q)^T{\Sigma_{\alpha}^q}^{-1}(\alpha-\mu_\alpha^q)) \\
   &=& \frac{p+q}{2}\log 2\pi+\frac{1}{2}\log |\Sigma_{\beta}^q|+\frac{1}{2}\log |\Sigma_{\alpha}^q| +\frac{p+q}{2}.
\eqan
In evaluating $T_2$ above we made use of the independence of $\beta$ and $\alpha$ in the variational
posterior and of the moment generating function for the multivariate normal variational 
posterior distribution for $\alpha$.  
Putting the terms together, the variational lower bound simplifies to (\ref{hetlowerbd}).
\section*{Appendix B}
Denote by $W(\a)$ the diagonal matrix $\diag(\frac12w_i\exp(-z_i^T\a))$, then
\beqn
u(\a):=\frac{\partial\log q(\a)}{\partial\a}=-\frac12\sum_iz_i+Z^TW(\a)-{\Sigma_\a^0}^{-1}(\a-\mu_\a^0)
\eeqn
and
\beqn
A(\a):=\frac{\partial^2\log q(\a)}{\partial\a\partial\a^T}=-Z^TW(\a)Z-{\Sigma_\a^0}^{-1}.
\eeqn
The Newton method for estimating the mode is as follows.
\begin{itemize}
\item Initialization: Set starting value $\a^{(0)}$.
\item Iteration: For $k=1,2,...$, update $\a^{(k)}=\a^{(k-1)}+A^{-1}(\a^{(k-1)})u(\a^{(k-1)})$ until some stopping rule is satisfied,
such as $\|\a^{(k)}-\a^{(k-1)}\|<\epsilon$ with some pre-specified tolerance $\epsilon$. 
\end{itemize}

\bibliographystyle{apalike}

\normalsize
\begin{thebibliography}{}

\bibitem[Aitchison, 1986]{Aitchison:1986}
Aitchison, J. (1986). 
\newblock{\em The Statistical Analysis of Compositional Data}, 
London: Chapman and Hall.

\bibitem[Beal and Ghahramani, 2003]{Beal:2003}
Beal, M. J., Ghahramani, Z. (2003).
The variational Bayesian EM algorithm for incomplete data: with application to scoring graphical model structures. 
{\em Bayesian Statistics}, 7, 453–-464.

\bibitem[Beal and Ghahramani, 2006]{Beal:2006}
Beal, M. J., Ghahramani, Z. (2006).
Variational Bayesian learning of directed graphical models with hidden variables. 
{\em Bayesian Analysis}, 1, 1–-44.

\bibitem[Bishop, 2006]{Bishop:2006}
Bishop, C. M. (2006). {\em Pattern Recognition and Machine Learning}. New York: Springer.

\bibitem[Brown et~al., 2001]{Brown:2001}
Brown, P. J., Fearn, T. and Vannucci, M. (2001).
Bayesian wavelet regression on curves with application to a spectroscopic calibration problem.
{\em Journal of the American Statistical Association}, 96, 398-408. 

\bibitem[Carroll and Ruppert, 1988]{Carroll:1988}
Carroll, R. J. and Ruppert, D. (1988). 
{\em Transformation and Weighting in Regression.} 
Monographs on Statistics and Applied Probability, Chapman and Hall, London.

\bibitem[Chan et~al., 2006]{Chan:2006}
Chan, D., Kohn, R., Nott, D. J. and Kirby, C. (2006). 
Adaptive nonparametric estimation of mean and variance functions. 
{\em Journal of Computational and Graphical Statistics}, 15, 915-936

\bibitem[Chen and Chen, 2008]{Chen:2008}
Chen, J. and Chen, Z. (2008).
Extended Bayesian information criteria for model selection with large model spaces.
{\em Biometrika}, 95, 759–-771.

\bibitem[Chib and Jeliazkov, 2001]{Chib:2001}
Chib, S. and Jeliazkov, I. (2001).
Marginal likelihood from the Metropolis-Hastings output.
{\em Journal of the American Statistical Association}, 96, 270-281.

\bibitem[Cottet et~al., 2008]{Cottet:2008}
Cottet, R., Kohn, R. and Nott, D.J. (2008).  
Variable selection and model averaging in overdispersed generalized linear models. 
{\em Journal of the American Statistical Association}, 103, 661-671.

\bibitem[Davidian and Carroll, 1987]{Davidian:1987}
Davidian, M. and Carroll, R. (1987).  Variance function estimation.  {\em Journal of the
American Statistical Association}, 82, 1079-1091.

\bibitem[Efron, 1986]{Efron:1986}
Efron, B. (1986).  Double exponential families and their use in generalised linear regression.
{\em Journal of the American Statistical Association}, 81, 709-721.

\bibitem[Efron et~al., 2004]{Efron:2004}
  Efron, B., Hastie, T., Johnstone, I., and Tibshirani, R. (2004).
  \newblock Least angle regression (with discussion).
  \newblock {\em The Annals of Statistics}, 32, 407--451.

\bibitem[Friedman, 2008]{Friedman:2008}
Friedman, J. H. (2008).
Fast sparse regression and classification. 2008. 
URL http://www-stat.stanford.edu/ jhf/ftp/GPSpaper.pdf

\bibitem[George and McCulloch, 1993]{George:1993}
George, E. I. and McCulloch, R. E. (1993).
Variable selection via Gibbs sampling.
{\em Journal of American Statistical Association}, 88, 881–-889.

\bibitem[Geweke and Keane, 2007]{Geweke:2007}
Geweke, J. and Keane, M. (2007). 
Smoothly mixing regressions. 
{\em Journal of Econometrics}, 138, 252--291.

\bibitem[Jordan et~al., 1999]{Jordan:1999}
Jordan, M. I., Ghahramani, Z., Jaakkola, T. S., Saul, L. K. (1999).  
An introduction to variational methods for graphical models. 
In M. I. Jordan (Ed.), Learning in Graphical Models. MIT Press, Cambridge.

\bibitem[Mallat and Zhang, 1993]{Mallat:1993}
Mallat, S. G. and Zhang, Z. (1993).
Matching pursuits with time-frequency dictionaries.
{\em IEEE Transactions on signal processing}, 41, 3397-3415.

\bibitem[Nelder and Pregibon, 1987]{Nelder:1987}
Nelder, J. and Pregibon, D. (1987). An extended quasi-likelihood function.
{\em Biometrika, 74}, 221-232.

\bibitem[O'Hagan and Forster, 2004]{O'Hagan:2004}
O'Hagan, A. and Forster, J. J. (2004). {\em Bayesian Inference}, 2nd edition, 
volume 2B of ``Kendall's Advanced Theory of Statistics". Arnold, London.
 
\bibitem[Ormerod and Wand, 2009]{Ormerod:2009}
Ormerod, J. T. and Wand, M. P. (2009).  Explaining variational approximation.  University of 
Wollongong technical report, available at \verb+http://www.uow.edu.au/~mwand/evapap.pdf+

\bibitem[Osborne et~al., 1984]{Osborne:1984}
Osborne, B. G., Fearn, T., Miller, A. R. and Douglas, S. (1984).
Application of near infrared reflectance spectroscopy to the
compositional analysis of biscuits and biscuit doughs.
{\em Journal of the Science of Food and Agriculture}, 35, 99-105.

\bibitem[Raftery et al., 1997]{Raftery:1997}
Raftery, A. E., Madigan, D. and Hoeting, J. (1997). 
Bayesian model averaging for linear regression models. 
{\em Journal of the American Statistical Association}, 92, 179-–191.

\bibitem[Rigby and Stasinopoulos, 2005]{Rigby:2005}
Rigby, R. A. and Stasinopoulos, D. M. (2005). 
Generalized additive models for location, scale and shape (with discussion).
{\em Applied Statistics}, 54, 507–-554.

\bibitem[Ruppert et~al., 2003]{Ruppert:2003}
Ruppert, D., Wand, M. P. and Carroll, R. J. (2003).
{\em Semiparametric regression}, Cambridge University Press.

\bibitem[Schapire, 1990]{Schapire:1990}
Schapire, R. E. (1990). 
The strength of weak learnability. 
{\em Machine Learning}, 5(2): 1997-2027.

\bibitem[Smith and Kohn, 1996]{Smith:1996}
Smith, M. and Kohn, R. (1996).
Nonparametric regression using Bayesian variable selection.
{\em Journal of Econometrics}, 75, 317-343.

\bibitem[Smyth, 1989]{Smyth:1989}
Smyth, G. (1989).
Generalized linear models with varying dispersion.
{\em Journal of the Royal Statistical Society}, B, 51, 47-60.

\bibitem[Tibshirani, 1996]{Tibshirani:1996}
  Tibshirani, R. (1996).
  \newblock Regression shrinkage and selection via the lasso.
  \newblock {\em Journal of the Royal Statistical Society, Series B}, 58, 
   267--288.

\bibitem[Tropp, 2004]{Tropp:2004}
Tropp, J. A. (2004).
Greed is good: algorithmic results for sparse approximation.
{\em IEEE Transactions on information theory}, 50, 2231-2242.

\bibitem[Villani et al., 2009]{Villani:2009}
Villani, M., Kohn, R. and Giordani, P. (2009). 
Regression density estimation using smooth adaptive Gaussian mixtures. 
{\em Journal of Econometrics}, 153, 155--173.

\bibitem[Weisberg, 2005]{Weisberg:2005}
Weisberg, S. (2005).  {\em Applied Linear Regression}, third edition, Hoboken NJ: John Wiley.

\bibitem[Wu et al., 2010]{Wu:2010}
Wu, B., McGrory, C. A. and Pettitt, A. N. (2010).
A new variational Bayesian algorithm with application to human mobility pattern modeling.
{\em Statistics and Computing}. DOI 10.1007/s11222-010-9217-9.

\bibitem[Yau and Kohn, 2003]{Yau:2003}
Yau, P. and Kohn, R. (2003).  Estimation and variable selection in nonparametric heteroscedastic regression.
{\it Statistics and Computing}, 13, 191-208. 

\bibitem[Yee and Wild, 1996]{Yee:1996}
Yee, T. and Wild, C. (1996).  Vector generalized additive models.  {\em Journal of the
Royal Statistical Society, Series B}, 58, 481-493.

\bibitem[Zhang, 2009]{Zhang:2009}
Zhang, T. (2009). 
On the consistency of feature selection using greedy least squares regression.
{\em Journal of Machine Learning Research}, 10, 555-568.

\bibitem[Zhao and Yu, 2007]{Zhao:2007}
Zhao, P. and Yu, B. (2007). Stagewise lasso. {\em Journal of Machine Learning Research}, 8, 2701–-2726.

\bibitem[Zou, 2006]{Zou:2006}
Zou, H. (2006). The adaptive Lasso and its oracle properties.
{\em Journal of the American Statistical Association}, 106, 1418--1429.
\end{thebibliography}
\renewcommand{\baselinestretch}{1}

\end{document}